\documentclass[acmsmall]{acmart}
\pdfoutput=1
\usepackage{booktabs}
\usepackage{multirow}
\usepackage{tikz}
\usepackage{graphicx}
\usepackage{pgfplots, pgfplotstable}
\usepackage{scalefnt}
\usepackage{makecell}

\usetikzlibrary{shapes.gates.logic.US,trees,positioning,arrows}
\usetikzlibrary{shapes.multipart}
\usetikzlibrary{shapes,arrows}
\usetikzlibrary{patterns}
\usepgfplotslibrary{statistics}
\pgfplotsset{compat=1.8}

\usepackage{graphicx}
\usepackage{amsmath}     
\usepackage{xcolor}
\usepackage{comment}
\usepackage{rotating}

\usepackage[final]{changes}
\definechangesauthor[name={Author 1}, color= violet]{author1}
\definechangesauthor[name={Author 2}, color=orange]{author2}
\definechangesauthor[name={Author 3}, color=olive]{author3}
\definechangesauthor[name={Author 4}, color=red]{author4}
\definechangesauthor[name={Author 5}, color=blue]{author5}

\AtBeginDocument{%
  \providecommand\BibTeX{{%
    \normalfont B\kern-0.5em{\scshape i\kern-0.25em b}\kern-0.8em\TeX}}}

\setcopyright{acmcopyright}
\copyrightyear{2020}
\acmYear{2020}
\acmDOI{XX.XXXX/XXXXXXX.XXXXXXX}

\acmJournal{JACM}
\acmVolume{XX}
\acmNumber{X}
\acmMonth{06}

\usepackage{caption}
\usepackage{subcaption}
\usetikzlibrary{spy,calc}
\usepackage{hyperref}

\newif\ifblackandwhitecycle
\gdef\patternnumber{0}

\pgfkeys{/tikz/.cd,
    zoombox paths/.style={
        draw=orange,
        very thick
    },
    black and white/.is choice,
    black and white/.default=static,
    black and white/static/.style={ 
        draw=white,   
        zoombox paths/.append style={
            draw=white,
            postaction={
                draw=black,
                loosely dashed
            }
        }
    },
    black and white/static/.code={
        \gdef\patternnumber{1}
    },
    black and white/cycle/.code={
        \blackandwhitecycletrue
        \gdef\patternnumber{1}
    },
    black and white pattern/.is choice,
    black and white pattern/0/.style={},
    black and white pattern/1/.style={    
            draw=white,
            postaction={
                draw=black,
                dash pattern=on 2pt off 2pt
            }
    },
    black and white pattern/2/.style={    
            draw=white,
            postaction={
                draw=black,
                dash pattern=on 4pt off 4pt
            }
    },
    black and white pattern/3/.style={    
            draw=white,
            postaction={
                draw=black,
                dash pattern=on 4pt off 4pt on 1pt off 4pt
            }
    },
    black and white pattern/4/.style={    
            draw=white,
            postaction={
                draw=black,
                dash pattern=on 4pt off 2pt on 2 pt off 2pt on 2 pt off 2pt
            }
    },
    zoomboxarray inner gap/.initial=5pt,
    zoomboxarray columns/.initial=2,
    zoomboxarray rows/.initial=2,
    subfigurename/.initial={},
    figurename/.initial={zoombox},
    zoomboxarray/.style={
        execute at begin picture={
            \begin{scope}[
                spy using outlines={%
                    zoombox paths,
                    width=\imagewidth / \pgfkeysvalueof{/tikz/zoomboxarray columns} - (\pgfkeysvalueof{/tikz/zoomboxarray columns} - 1) / \pgfkeysvalueof{/tikz/zoomboxarray columns} * \pgfkeysvalueof{/tikz/zoomboxarray inner gap} -\pgflinewidth,
                    height=\imageheight / \pgfkeysvalueof{/tikz/zoomboxarray rows} - (\pgfkeysvalueof{/tikz/zoomboxarray rows} - 1) / \pgfkeysvalueof{/tikz/zoomboxarray rows} * \pgfkeysvalueof{/tikz/zoomboxarray inner gap}-\pgflinewidth,
                    magnification=3,
                    every spy on node/.style={
                        zoombox paths
                    },
                    every spy in node/.style={
                        zoombox paths
                    }
                }
            ]
        },
        execute at end picture={
            \end{scope}
            \node at (image.north) [anchor=north,inner sep=0pt] {\subcaptionbox{\label{\pgfkeysvalueof{/tikz/figurename}-image}}{\phantomimage}};
            \node at (zoomboxes container.north) [anchor=north,inner sep=0pt] {\subcaptionbox{\label{\pgfkeysvalueof{/tikz/figurename}-zoom}}{\phantomimage}};
     \gdef\patternnumber{0}
        },
        spymargin/.initial=0.5em,
        zoomboxes xshift/.initial=1,
        zoomboxes right/.code=\pgfkeys{/tikz/zoomboxes xshift=1},
        zoomboxes left/.code=\pgfkeys{/tikz/zoomboxes xshift=-1},
        zoomboxes yshift/.initial=0,
        zoomboxes above/.code={
            \pgfkeys{/tikz/zoomboxes yshift=1},
            \pgfkeys{/tikz/zoomboxes xshift=0}
        },
        zoomboxes below/.code={
            \pgfkeys{/tikz/zoomboxes yshift=-1},
            \pgfkeys{/tikz/zoomboxes xshift=0}
        },
        caption margin/.initial=4ex,
    },
    adjust caption spacing/.code={},
    image container/.style={
        inner sep=0pt,
        at=(image.north),
        anchor=north,
        adjust caption spacing
    },
    zoomboxes container/.style={
        inner sep=0pt,
        at=(image.north),
        anchor=north,
        name=zoomboxes container,
        xshift=\pgfkeysvalueof{/tikz/zoomboxes xshift}*(\imagewidth+\pgfkeysvalueof{/tikz/spymargin}),
        yshift=\pgfkeysvalueof{/tikz/zoomboxes yshift}*(\imageheight+\pgfkeysvalueof{/tikz/spymargin}+\pgfkeysvalueof{/tikz/caption margin}),
        adjust caption spacing
    },
    calculate dimensions/.code={
        \pgfpointdiff{\pgfpointanchor{image}{south west} }{\pgfpointanchor{image}{north east} }
        \pgfgetlastxy{\imagewidth}{\imageheight}
        \global\let\imagewidth=\imagewidth
        \global\let\imageheight=\imageheight
        \gdef\columncount{1}
        \gdef\rowcount{1}
        
    },
    image node/.style={
        inner sep=0pt,
        name=image,
        anchor=south west,
        append after command={
            [calculate dimensions]
            node [image container,subfigurename=\pgfkeysvalueof{/tikz/figurename}-image] {\phantomimage}
            node [zoomboxes container,subfigurename=\pgfkeysvalueof{/tikz/figurename}-zoom] {\phantomimage}
        }
    },
    color code/.style={
        zoombox paths/.append style={draw=#1}
    },
    connect zoomboxes/.style={
    spy connection path={\draw[draw=none,zoombox paths] (tikzspyonnode) -- (tikzspyinnode);}
    },
    help grid code/.code={
        \begin{scope}[
                x={(image.south east)},
                y={(image.north west)},
                font=\footnotesize,
                help lines,
                overlay
            ]
            \foreach \x in {0,1,...,9} { 
                \draw(\x/10,0) -- (\x/10,1);
                \node [anchor=north] at (\x/10,0) {0.\x};
            }
            \foreach \y in {0,1,...,9} {
                \draw(0,\y/10) -- (1,\y/10);                        \node [anchor=east] at (0,\y/10) {0.\y};
            }
        \end{scope}    
    },
    help grid/.style={
        append after command={
            [help grid code]
        }
    },
}

\newcommand\phantomimage{%
    \phantom{%
        \rule{\imagewidth}{\imageheight}%
    }%
}
\newcommand\zoombox[2][]{
    \begin{scope}[zoombox paths]
        \pgfmathsetmacro\xpos{
            (\columncount-1)*(\imagewidth / \pgfkeysvalueof{/tikz/zoomboxarray columns} + \pgfkeysvalueof{/tikz/zoomboxarray inner gap} / \pgfkeysvalueof{/tikz/zoomboxarray columns} ) + \pgflinewidth
        }
        \pgfmathsetmacro\ypos{
            (\rowcount-1)*( \imageheight / \pgfkeysvalueof{/tikz/zoomboxarray rows} + \pgfkeysvalueof{/tikz/zoomboxarray inner gap} / \pgfkeysvalueof{/tikz/zoomboxarray rows} ) + 0.5*\pgflinewidth
        }
        \edef\dospy{\noexpand\spy [
            #1,
            zoombox paths/.append style={
                black and white pattern=\patternnumber
            },
            every spy on node/.append style={#1},
            x=\imagewidth,
            y=\imageheight
        ] on (#2) in node [anchor=north west] at ($(zoomboxes container.north west)+(\xpos pt,-\ypos pt)$);}
        \dospy
        \pgfmathtruncatemacro\pgfmathresult{ifthenelse(\columncount==\pgfkeysvalueof{/tikz/zoomboxarray columns},\rowcount+1,\rowcount)}
        \global\let\rowcount=\pgfmathresult
        \pgfmathtruncatemacro\pgfmathresult{ifthenelse(\columncount==\pgfkeysvalueof{/tikz/zoomboxarray columns},1,\columncount+1)}
        \global\let\columncount=\pgfmathresult
        \ifblackandwhitecycle
            \pgfmathtruncatemacro{\newpatternnumber}{\patternnumber+1}
            \global\edef\patternnumber{\newpatternnumber}
        \fi
    \end{scope}
}

\begin{document}

\title[Designing for Recommending Intermediate States in A SWfMS]{Designing for Recommending Intermediate States in A Scientific Workflow Management System}

\author{Debasish Chakroborti}
\affiliation{%
  \institution{Department of Computer Science, University of Saskatchewan}
  \streetaddress{176 Thorvaldson Bldg, 110 Science Place}
  \city{Saskatoon}
  \state{Saskatchewan}
  \postcode{S7N 5C9}
  \country{Canada}}
\email{debasish.chakroborti@usask.ca}
\orcid{0000-0002-1597-8162}
\author{Banani Roy}
\affiliation{%
  \institution{Department of Computer Science, University of Saskatchewan}
  \streetaddress{176 Thorvaldson Bldg, 110 Science Place}
  \city{Saskatoon}
  \state{Saskatchewan}
  \postcode{S7N 5C9}
  \country{Canada}}
\email{banani.roy@usask.ca}
\author{Sristy Sumana Nath}
\affiliation{%
  \institution{Department of Computer Science, University of Saskatchewan}
  \streetaddress{176 Thorvaldson Bldg, 110 Science Place}
  \city{Saskatoon}
  \state{Saskatchewan}
  \postcode{S7N 5C9}
  \country{Canada}}
\email{sristy.sumana@usask.ca}
%

\renewcommand{\shortauthors}{D. Chakroborti et al.}

\begin{abstract}
To process a large amount of data sequentially and systematically,  proper management of workflow components (i.e., modules, data, configurations, associations among ports and links) in a Scientific Workflow Management System (SWfMS) is inevitable. Managing data with provenance in a SWfMS to support reusability of workflows, modules, and data is not a simple task.  Handling such components is even more burdensome for frequently assembled and executed complex workflows for investigating large datasets with different technologies (i.e., various learning algorithms or models). However, a great many studies propose various techniques and technologies for managing and recommending services in a SWfMS, but only a very few studies consider the management of data in a SWfMS for efficient storing and facilitating workflow executions. Furthermore, there is no study to inquire about the effectiveness and efficiency of such data management in a SWfMS from a user perspective. In this paper, we present and evaluate a GUI version of such a novel approach of intermediate data management with two use cases (Plant Phenotyping and Bioinformatics).  The technique we call GUI-RISP\textsubscript{TS} (Recommending Intermediate States from Pipelines Considering Tool-States) can facilitate executions of workflows with processed data (i.e., intermediate outcomes of modules in a workflow) and can thus reduce the computational time of some modules in a SWfMS. We integrated GUI-RISP\textsubscript{TS} with an existing workflow management system called SciWorCS.  In SciWorCS, we present an interface that users use for selecting the recommendation of intermediate states (i.e., modules' outcomes). We investigated GUI-RISP\textsubscript{TS}'s effectiveness from users' perspectives along with measuring its overhead in terms of storage and efficiency in workflow execution.
\end{abstract}




\maketitle

\section{Introduction}

In Big-data analytics when a large volume of heterogeneous data needs to be processed with different technologies (i.e., various learning algorithms or models), proper data flow and module/service associations (a.k.a workflows) in a SWfMS are crucial for efficiency. Users often assemble workflows manually in SWfMSs, where processing modules are selected for the purpose of a particular task. Besides, users frequently run similar workflows by changing only a few modules for finding suitable methodologies to investigate on a particular input dataset \cite{inproceedingsBigD18}. In such frequently executed workflows, processing of large datasets eventually requires some computationally expensive modules. Investigating the same dataset with several workflows by comprising the computationally expensive modules ultimately requires a long execution time \cite{Chakroborti2020}. In such a situation, to experiment with new methodologies by changing only a few modules of a workflow could be a tedious task. Since modules are reusable in different workflows, similarly consideration of reusable outcomes of the modules in a SWfMS could be a solution for increasing the performance of a workflow execution.

\replaced[id=author1, remark={Addressed for the reviews of Round 2 Reviewer 3}]{
However, improving the performance of a workflow by introducing reusability and storing all of the generated data in a SWfMS is not a feasible solution as the generated data or intermediate data are also large enough to increase the storing and storage cost dramatically. As well as storing all of the generated data from a workflow can hamper the resource utilization (i.e., occupying a large portion of persistent storage) in a SWfMS. In a workflow assembling process, where different ports (i.e., inputs and outputs of a module) and parameter configurations for different types of data and data flow are involved, a proper decision-making scheme is necessary in the runtime to store data, to support reusability, to introduce efficiency and to manage storage cost of the heterogeneous data. Consequently, proper management of such data and metadata (i.e., association information of module to data, module to module and module to configuration) in a system of workflow processing is essential to facilitate the execution of workflows at low costs.}{}

\replaced[id=author1, remark={Addressed for the reviews of Round 2 Reviewer 3}]{
To introduce such data management in a previous work, Chakroborti et. al \cite{inproceedingsBigD18} proposed a technique of automatic data recommendation for storing and retrieving intermediate data of workflows in a SWfMS. 
The technique, known as RISP (Recommending Intermediate  States from Pipelines), does not consider parameter configuration of modules while recommending intermediate states. The main problem with the technique is that it might suggest unwanted intermediate data to users. Another problem is that users might not get all the options that can be produced at module outcomes for different configuration settings. A technique is required where parameter configuration will be utilized for mitigating the problems. The proposed technique in this study with a data management scheme is intended to facilitate workflow assembling and executions with different configuration sets. Furthermore, this study is also designed to evaluate the technique in real-world SWfMS from users' perspectives.}{}

\replaced[id=author1, remark={Addressed for the reviews of Round 2 Reviewer 3}]{A GUI (Graphical User Interface) version of the technique is considered in SciWorCS (Collaborative Scientific Workflow Management System) \cite{10.1145/3331151} system - a groupware system for scientific data analysis, where users can have more control and choice to select intermediate data. Multiple intermediate datasets in a SWfMS can be found for different states of a corresponding tool. Such as parameter configurations or tunings in a tool can generate different intermediate states in different workflows. If we consider the tool states in the association rules of the RISP by considering parameter matching, we can give more choices to users for selecting intermediate states. In this paper we call our proposed approach GUI-RISP\textsubscript{TS} meaning Graphical User Interface for Recommending Intermediate States from Pipelines Considering Tool-States. More details of this GUI version have been discussed in Section \ref{technique_integration_overview}.}{}

Nowadays, SWfMSs are used to manage modules and data in a way that can facilitate building workflows interactively for reducing configuration hassles \cite{article34636} \cite{4404805}  \cite{4534284} \cite{Simmhan:2005:SDP:1084805.1084812} \cite{10.1007/11890850_2} \cite{Gray:2005:SDM:1107499.1107503} \cite{Davidson:2008:PSW:1376616.1376772}. Modules are usually local scripts or web-services and most of the SWfMSs support interactive drag and drop facilities to add such modules while building workflows from a set of available modular tools. For a particular task, port linking, dataset importing, and parameter configuring are also possible interactively to assemble sequential execution or build workflows as an acyclic graph in a SWfMS. Besides, a SWfMS also coordinates processing scripts, datasets and resource allocations in various steps of workflow execution to ensure dependencies of dataflow. For such systems, quantitative analysis of performance is hard to measure, and only organizational performance (i.e., performances related to user experiences) by the user evaluation can be interpreted with use cases \cite{REIJERS2016126} \cite{772961}. Also, the effectiveness of a SWfMS and integration of a new methodology in the SWfMS to increase efficiency could be assessed by user evaluation. Thus, in this study, we evaluate our intermediate data recommendation technique in the SciWorCS system and assess the feasibilities in various aspects of assembling and executing workflows.

\added[id=author2, remark={Addressed for the reviews of Reviewer 3 and 4}]{
The overarching research question that the paper addresses is how we can handle intermediate states optimally in SWfMSs considering users' choices? In order to get the answer of this question, we developed GUI-RISP\textsubscript{TS}. In GUI-RISP\textsubscript{TS} we first developed an algorithm based on association rules and workflow tools' configuration settings or states to store intermediate states optimally (discussed in Section \ref{technique_integration_overview}). Then we introduce the technique with a graphical user interface (GUI) in the SciWorCS that helps users handle intermediates states intuitively.  
We present the contribution of our work in the form of the following five research questions.}

\begin{itemize}

\item \textbf{RQ1:} \added[id=author2, remark={Addressed for the reviews of Reviewer 3 and 4}]{How much performance overhead GUI-RISP\textsubscript{TS} adds in the system?} 

\added[id=author2, remark={Addressed for the reviews of Reviewer 3 and 4}]{To measure the performance overhead of GUI-RISP\textsubscript{TS} we did both system-level and user-level performance tests using Apache Jmeter \cite{halili2008apache} and Apdex Standard \cite{sevcik2005defining} respectively.}

\item \textbf{RQ2:} \added[id=author2, remark={Addressed for the reviews of Reviewer 3 and 4}]{How can we incorporate tools' states in recommending intermediate states using association rules?}

\added[id=author2, remark={Addressed for the reviews of Reviewer 3 and 4}]{We extended the algorithm proposed by Chakroborti et al. \cite{inproceedingsBigD18} for considering the tool states. As the prior approach did not consider tool states, it could recommend only one intermediate state for a particular tool execution. Whereas our approach can store multiple intermediate states for a particular tool and gives options to the user to select an intermediate state based on their requirement.}

\item \textbf{RQ3:} \added[id=author2, remark={Addressed for the reviews of Reviewer 3 and 4}]{Can we compose workflows while specifying intermediate states?} 

\added[id=author2, remark={Addressed for the reviews of Reviewer 3 and 4}]{To answer this question, we designed a GUI on top of SciWorCS and described how users would interact with the different elements of the GUI. We consider two domains - image processing and bioinformatics for our case studies and various experiments with the GUI.}

\item \textbf{RQ4:} \added[id=author2, remark={Addressed for the reviews of Reviewer 3 and 4}]{How do users perceive intermediate states in composing workflows?} 

\added[id=author2, remark={Addressed for the reviews of Reviewer 3 and 4}]{We conducted user study on two different settings: data-intensive and data-nonintensive to figure out how users use intermediate states.}

\item \textbf{RQ5:} \added[id=author2, remark={Addressed for the reviews of Reviewer 3 and 4}]{What are the major areas where participants are mostly involved while assembling and executing workflows with the GUI-RISP\textsubscript{TS}?}

\added[id=author2, remark={Addressed for the reviews of Reviewer 3 and 4}]{To distinguish between the key areas, we presented recommendations and composition areas in different area-of-interest for being used by an eye-tracker.} 

\end{itemize}
\par

\added[id=author2, remark={Addressed for the reviews of Reviewer 3 and 4}]{The study is designed to demonstrate the integration synopsis of the GUI-RISP\textsubscript{TS} in SciWorCS and evaluate the SWfMS with the data recommendation technique from users usage perspectives by executing various types of workflows. This study is also outlined in different subsections to answer the RQs that provide us the following findings:}

\begin{itemize}
  
  \item  \added[id=author2, remark={Addressed for the reviews of Reviewer 3 and 4}]{By analyzing workflow assembling and executing log from our user study it can be illustrated that the GUI-RISP\textsubscript{TS} can help users in executing pipelines efficiently (e.g., 56\% less user request, and 25\% less time in workflow execution).}

  \item  \added[id=author2, remark={Addressed for the reviews of Reviewer 3 and 4}]{GUI-RISP\textsubscript{TS} can be more effective while developing long, complex workflows (e. g., 50\% more use in intermediate data than the short workflows from the available intermediate data).}

  \item  \added[id=author2, remark={Addressed for the reviews of Reviewer 3 and 4}]{GUI-RISP\textsubscript{TS} in SciWorCS can provide adequate information to users for designing workflows with intermediate data and getting control over data on the interface.}

\end{itemize}
Our findings indicate that our proposed technique can significantly improve the performance and efficiency of a SWfMS by reusing the intermediate states from previously executed workflows.

The rest of the paper is organized as follows. Section \ref{RelatedWorks} discusses the related work, Section \ref{technique_integration_overview} presents architecture of our proposed technique, Section \ref{background} defines and describes some basic terms related to the study, 
Section \ref{ImplementationDetails} describes our experimental setup, 
Section \ref{ExperimentalStudiesandResults} presents the user evaluations, Section \ref{discussion_and_future_direction} shows some future directions,
 Section \ref{ThreattoValidity} describes possible threats to validity
and finally, Section \ref{ConclusionandFutureWorks} concludes the paper by mentioning our future directions.

\section{Related Work}
\label{RelatedWorks}

Most of the SWfMSs are implemented as a process-aware based information system where data management is mostly neglected in terms of reuse for facilitating workflow execution. In this paper, our study is mainly focused on user evaluation of a data recommendation and management technique in a SWfMS. Existing studies on recommendation, management and new technique evaluation for both data and processes in SWfMSs are presented below in two sections to compare with our proposed technique.

\subsection{Recommendations in SWfMSs}

A number of studies \cite{Woodman2015WorkflowProvenance} \cite{Yuan2011On-demandSystems} \cite{Koop2008VisComplete:Pipelines} \cite{Gil:2011:MYM:2063076.2063082} \cite{10.1007/978-3-540-85502-6_18} have been conducted on recommending and managing modules and data in  SWfMSs. Woodman et al. \cite{Woodman2015WorkflowProvenance} proposed an algorithm for determining which subset of the intermediate state results from a pipeline could be stored at the lowest cost. Similarly, Yuan et al. \cite{Yuan2011On-demandSystems} proposed an algorithm for determining which set of intermediate state results could be stored at a minimum cost from a SWfMS.
\added[id=author1, remark={Addressed for the reviews of Round 2 Reviewer 3}]{Both of the algorithms are sufficient to introduce a cost-effective way of storing generated data, but not suitable for introducing reusability of the generated data in a SWfMS.} Koop et al. \cite{Koop2008VisComplete:Pipelines} proposed a technique for automatically completing a pipeline under progress by analyzing pipeline execution history. Gil et al. \cite{Gil:2011:MYM:2063076.2063082} used metadata in their technique to select an appropriate model for a particular dataset and to set up modules' parameters dynamically. Likewise, Leake et al. \cite{10.1007/978-3-540-85502-6_18} demonstrated the use of semantic information of provenance for suggesting services. Besides, many studies \cite{Chinthaka2009CBRAssistant} \cite{Zhang2011Recommend-as-you-go:Reuse} \cite{Spjuth2015ExperiencesBioinformatics} have investigated the possibility of discovering and reusing services and workflows in a SWfMS. 
\added[id=author1, remark={Addressed for the reviews of Reviewer 4}]{Chakroborti et. al \cite{inproceedingsBigD18} proposed a technique of automatic data recommendation for storing and retrieving intermediate data of workflows in a SWfMS. Although the technique introduces data reusability, the recommendation could be unwanted for not considering the tool states (i.e., a user might not want a recommendation of outcome from different parameter configurations, even though the tool can not provide the configuration set information).}
While the above studies solely focus on storage cost of data or module recommendation, our study is fundamentally different because we focus on an accurate re-usability technique by recommending the storing of the outcomes of modules with parameter configurations and in this study, we also evaluate the effectiveness of the technique.
From the above inspection, it is also clear that only a very few of the existing studies investigated on intermediate state results for reuse in a SWfMS. In this study, we present the details architecture and user evaluation of our recommendation technique of intermediate data storing and reusing in a SWfMS. Additionally, use cases from different areas of workflow building are presented to evaluate the proposed technique.
\subsection{Case Studies, Technology Integrations and Evaluations}

A number of studies \cite{REIJERS2016126} \cite{772961} have been done on Use-case analysis of system integration and architecture implementation in the domain of Scientific Workflow Management. Some of them are discussed here to understand the effectiveness and evaluation process of a SWfMS. Wu et al. \cite{wu2012} illustrated a distributed WfMS's implementation with web services and emphasized on the quality of mapping schemes of service composing algorithms for performance enhancement. Zheng et al. \cite{Zheng:2015:ICW:2755979.2755984} analyzed the effects of Docker container for workflows to make them compatible in various infrastructures and explored the possibilities of virtual container management for low overhead in a SWfMS. Muniswamy-Reddy et al. \cite{Muniswamy-Reddy:2006:PSS:1267359.1267363} proposed a storage system by considering low overhead in transactions where provenance traces are considered as metadata to provide additional functionalities of a typical file system. Brown et al. \cite{Brown2007}  analyzed LIGO WfMS by integrating techniques and technologies from various WfMSs. In their study of LIGO system, gravitational wave data are used to explore future directions and usefulness of the techniques of different WfMSs in a single system. Wang et al. \cite{Wang:2009:KHG:1645164.1645176} demonstrated the architecture of the distributed Kepler and explored its performance and future directions for data-intensive tasks. Bahsi et al. \cite{Bahsi:2007:CWM:1377549.1377550} have analyzed different SWfMSs from the perspective of conditional-workflow-building process and stated that conditions could be managed in various ways in different SWfMSs respect to their core implementations. Gathering the facts from the above studies on evaluating and verifying the effectiveness of SWfMSs, in this study, we also want to investigate the use cases of our technique in a SWfMS.

\added[id=author1, remark={Addressed for the reviews of Reviewer 3}]{By focusing on the limitation of the existing works, our study is presented to propose a modified version of the reusable data recommendation technique. The previous data recommendation technique is based on the tool IDs. But, our study considers the state of the tools in a SWfMS and their IDs while recommending reusable data. Details of the preference of tool states are discussed in Section \ref{technique_integration_overview}.} It should be noted that recent studies that handle large datasets for pipeline based executions (i.e., TensorFlow Dataset \footnote{https://www.tensorflow.org/datasets}) are usually introduced for unified data gathering mechanisms. Our technique GUI-RISP\textsubscript{TS} works in the runtime of a pipeline after getting data from such a unified input data model.

\section{GUI-RISP\textsubscript{TS} in a SWfMS}
\label{technique_integration_overview}

\begin{figure}
  \includegraphics[width=14cm, height=16cm, keepaspectratio]{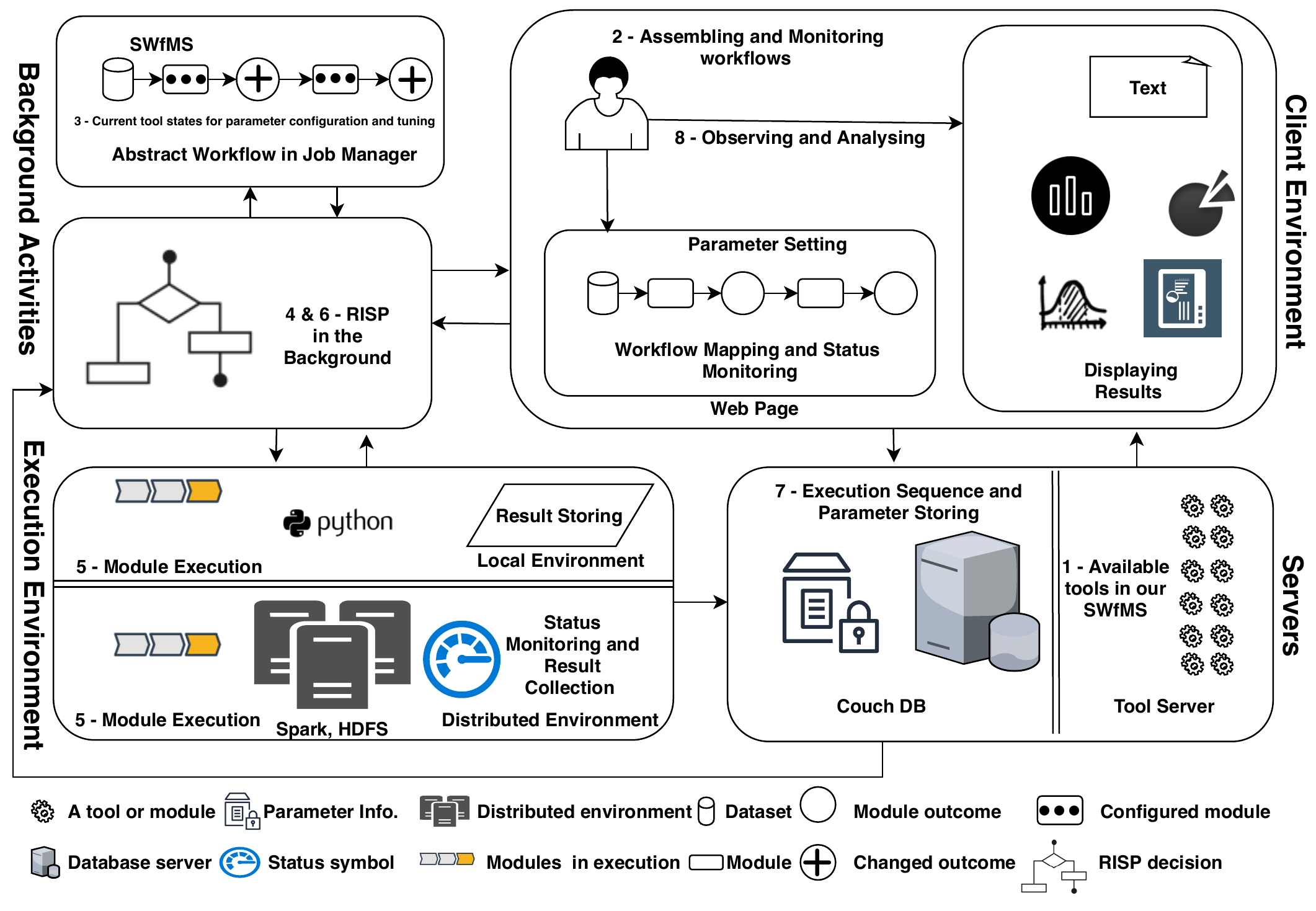}
  \caption{High level architecture of the SWfMS with GUI-RISP\textsubscript{TS}}
  \label{RISPArchitecture}  
\end{figure}

We present a brief overview of SciWorCS architecture with the integrated GUI version of RISP (GUI-RISP\textsubscript{TS} - A tool to assemble workflows with recommended reusable intermediate data). 
By Integrating GUI-RISP\textsubscript{TS} in the SWfMS, scientific analysis can be done by a set of reusable modules and intermediate data. Intermediate data could be preserved in SciWorCS by different users from previously executed workflows. In SciWorCS, workflows are assembled on a web interface where modules could be executed in both distributed and local environments by their implementation. Similarly, datasets could be stored in a local file system or a distributed file system (i.e., HDFS) for their purposes. RISP mainly works on background by getting some control information from the web-based GUI. Figure \ref{RISPArchitecture} illustrates the high-level overview of the architecture and the GUI-RISP\textsubscript{TS} integration of SciWorCS. In the figure, major components and user activities are numbered from 1 to 8 for clarifying the data flow in SciWorCS while building a workflow by using the GUI-RISP\textsubscript{TS}. \added[id=author1, remark={Addressed for the reviews of Reviewer 1}]{In the figure, at the bottom-right corner tool server is presented with tool symbols. Available tools are preconfigured with default configuration sets. The data server is presented beside the tool server, where execution logs are stored after every successful workflow completion. Clients can assemble, execute, and monitor workflows in the client environment, presented at the top-right portion of the figure. A parameter setting with a different configuration set can occur here for various modules. Here monitoring means checking the status and comprehending the displayed results. The top-left side is presented with an abstract workflow. The first data symbol illustrates a data source, the second rectangle is a process, and the third circle is an outcome in the workflow. Dots in the process present the configuration sets of the modules in the workflow. The plus sign in the outcomes represents variations in the outcomes for different configuration sets. There is another background activity just below the abstract workflow, which is our proposed technique GUI-RISP\textsubscript{TS} for making runtime recommendation of intermediate data. The bottom-left of the figure is shown with the execution environment of the scientific workflow management system. Module execution can happen both in Python and distributed environments. This execution environment always communicates with the database-server and web-server for status checking and data storing}.
\subsection{System Overview}
\subsubsection{Reusable Module and Data}

Workflows for scientific data analysis are often considered as Directed Acyclic Graphs (DAGs) \cite{GARG2015256} \cite{Prodan:2005:DSS:1066677.1066835}. By considering intermediate data, a workflow can be presented as $W = (D, M, E, ID, O)$, where $M$ is a set of process modules with $n$ elements $m_i, (1 \leq i \leq n)$ and $E$ is set of directed edges of data flow $e_{ij} = (m_i, m_j), (1 \leq i \leq n, 1 \leq j \leq n, i \neq j)$ denoting the dependency and condition among modules and datasets. $D$ is the input data set, $O$ is the output data set, and $ID$ represents the set of outcomes of different modules (i.e., intermediate data). Intermediate data sets (IDs) from various workflows are considered to be managed with GUI-RISP\textsubscript{TS} for their incremental storage consumption.  A module from a set M especially performs a specific task (i. e., data preparation or feature extraction or model fitting or statistical analysis) in the SWfMS (i.e., SciWorCS 
). Computation scenario of such a module can be further customized with parameter settings of a specific configuration set $C$ -  where parameter configuration set is $P$, $c_i \epsilon C, (1 \leq  i \leq P)$.  By using a specific parameter set $P$, a module of set M might have in a different state from a state of the same module in similar or dissimilar workflows. Hence a workflow module and intermediate data can be generalized by $Definition$ \ref{def_module} and $Definition$ \ref{def_data} in SciWorCS respectively.

\begin{definition}
\textit{Process module is defined as a tuple, $m \Rightarrow \langle id, I, O,$ $ C, S, T, Id \rangle $ where $id$ is the unique identifier of the module in a system, $I$ and $O$ supported input and output formats. $C$ is the configuration set of different parameters, $S$ is the source code, $T$ is the state of the module in workflow execution, and the $Id$ is a set of intermediate data generated from a module in workflow execution.}
\label{def_module}
\end{definition}

\replaced[id=author1, remark={Addressed for the reviews of Round 2 Reviewer 2}]{
It should be noted that the state of a module depends on parameter configurations. Anyone anytime can change the parameter configuration and the state matters only when workflow execution begins. Workflow execution also set the configuration set final in our system. So, the state T is from beginning to end of workflow execution. The generalized definition can be helpful to readers for understanding the reusable modules of the SWfMS while building workflows.}{}

\begin{definition}
\textit{Intermediate dataset is defined as a tuple, $Id \Rightarrow \langle Sid, Did, S,$ $ sT, lT, m, T  \rangle $ where $Sid$ is the unique identifier of the dataset in a system, $Did$ raw dataset id. $S$ is the size of the datasets, $sT$ and $lT$ are the required saving and loading time for the dataset, $m$ is the module that generated the dataset with state $T$.}
\label{def_data}
\end{definition}

\replaced[id=author1, remark={Addressed for the reviews of Round 2 Reviewer 2}]{
It also should be noted that our recommendation is dataset dependent, so for a particular execution in a particular environment, the loading and saving time is independent and unique for a particular configuration. The generalized definition can be helpful to readers for understanding the reusable intermediate data of the SWfMS while building workflows.}{}

\subsubsection{Workflow Composition with Intermediate Data}
\begin{figure}
  \includegraphics[width=1\textwidth]{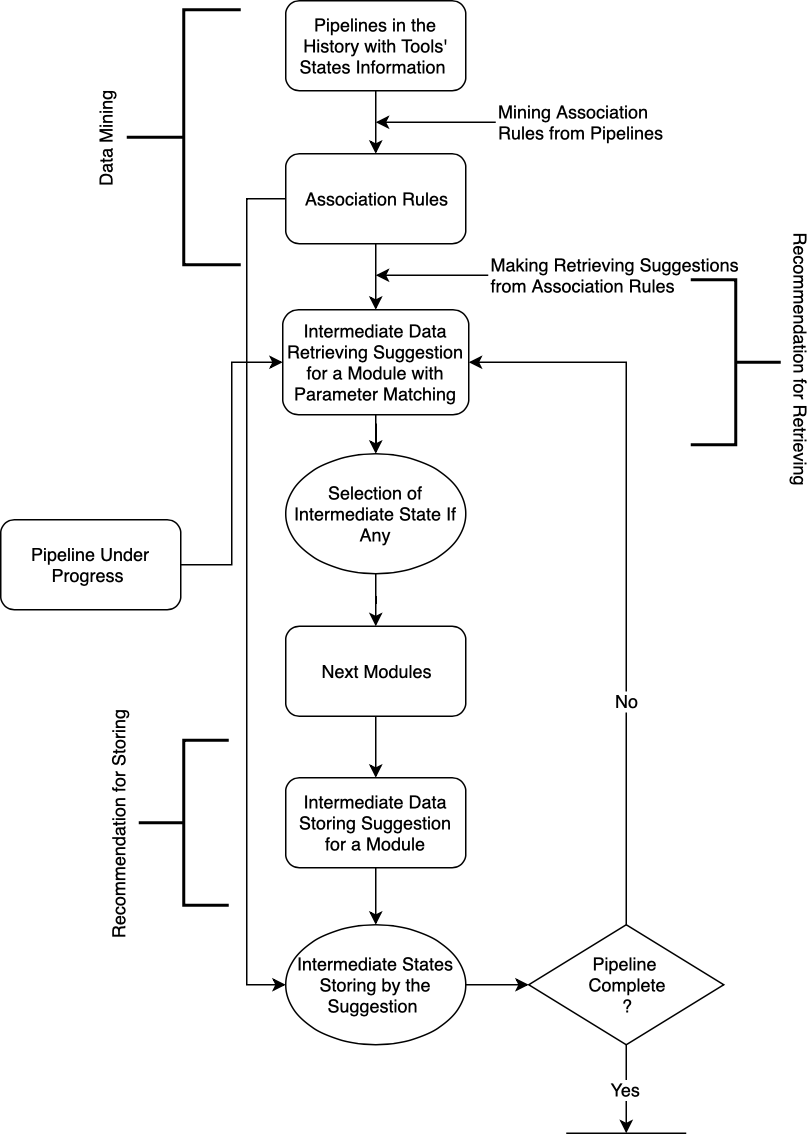}
  \caption{Workflow composition with the proposed technique }
  \label{IDRecommendationSystem}
\end{figure}

To perform data analysis with a workflow $W = (M, E)$, modules, $M$, from the toolbox of web panel need to be selected and assembled in the main composition window of SciWorCS. Besides, ports need to be mapped with links for the set $E$ to sustain the required dependencies of the workflow. Simply executing all of the modules from the workflow in a SWfMS would cost a considerable amount of time. For example, any given instance of the workflow execution, $W_i$, there are some modules outcomes $(Ids \epsilon ID)$ and a set $Id$ passed from one module to another for a dependency - defined by a directed edge, $e_{ij} = (m_i, m_j) \epsilon E$, and each module with its state $T$ perform some operations on the outcomes of the previous module $(Id_i)$ to produce some new outcomes $(Id_j)$. Usually, to perform operations on a large dataset, a single module takes a considerable amount of time, suppose $t_i$ time to process and generate an intermediate data $(Id_j)$. In most of the cases, these generated Id sets are also large and incrementally produced in a SWfMS. Since a workflow could be composed with a considerable number of modules, the execution time of the workflow might be too long. Again in a SWfMS, users generally build similar workflows frequently by changing only a few modules, i.e., $(m_k \dots m_l) \epsilon M $. In such a situation, rather than executing all modules $(m_1 \dots m_n) \epsilon M $ for a new workflow, in a SWfMS previously executed results ($IDs$) could be used to reduce execution cost. Besides, consideration of the state of a tool for derived data could increase user choice in a SWfMS. 
Hence, we adopt the GUI-RISP\textsubscript{TS} in  SciWorCS for each possible outcome of a module to consider the state of modules. By using the GUI version, we can recommend to reuse existing appropriate intermediate data while building new workflows in a SWfMS with proper parameter matching and data generation time information. 
Figure \ref{IDRecommendationSystem} demonstrates the core architecture of our data management and recommendation technique of GUI-RISP\textsubscript{TS} in SciWorCS.  

\subsubsection{Parameter Matching in a Composition}

\replaced[id=author1, remark={Addressed for the reviews of Reviewer 2}]{Suppose a module sequence $(m_1 \dots m_k) \epsilon M $ for a new workflow matched with a sequence of a previously executed workflow. However, a computation scenario of the previous workflow could customize the parameter setting of a specific parameter configuration set $P_{old}$ of the module sequence. The new workflow's unchanged sequence is also might be configured with a parameter set $P_{new}$. In such a case, users will get notification of recommendation from  GUI-RISP\textsubscript{TS} if the sets, $P_{old}$ and $P_{new}$ are identical with default or changed set of the similar module sequence.  However, if the data were not stored for the sequence, $(m_1 \dots m_k)$, for previous support and confidence value, and if new confidence value shows that the sequence is frequent now, then the intermediate data for the sequence, $(m_1 \dots m_k)$, are stored with the parameter set $P_{new}$. In future, if the intermediate data is generated with a different parameter set $P_{new2}$, it will be stored uniquely for the same sequence if the sequence is frequent at that time. This way, the same sequence, $(m_1 \dots m_k)$, can be recommended with multiple suggestions. Multiple suggestions are presented with the percentage of parameter matching. Percentages have come from the no. of the matched and available parameters for a sequence. Figure \ref{IDRecommendationSystem} shows that pipeline execution history with tool states information is used to recommend data retrieving suggestions for the pipeline under progress. Here, association rules are also generated with parameter information. In the figure, we can see that both past and current workflow execution histories are used for RISP recommendation. We used the same formula of association rules that are discussed in \cite{inproceedingsBigD18}, but rules are now distinct with their parameter configuration sets. Details of the association rules, their support, and confidence are discussed in Section \ref{background}. So, our \textbf{RQ2} can be answered that we can consider tool states in RISP's recommendation, and it can give us multiple suggestions with accurate information.}{}

\subsection{Background}
\label{background}

Association rules have been used in many research areas for discovering correlations and can be defined in the following way. 

\textbf{Association Rule.}
An association rule \cite{Agrawal:1993:MAR:170036.170072} is an expression of the form $X => Y$ where $X$ is the antecedent, and $Y$ is the consequent. Each of $X$ and $Y$ is a set of one or more program entities. 
The meaning of such a rule is that if $X$ gets changed in a particular commit operation, $Y$ also has the tendency of being changed in that commit.

\textbf{Support and Confidence.}
As defined by Zimmermann et al. \cite{Zimmermann:2004:MVH:998675.999460}, \emph{support is the number of commits in which an entity or a group of entities changed together}.
The support of an association rule is determined in the following way.

\begin{equation}
    \mathit{support}(X=>Y)=\mathit{support}(X, Y)
\end{equation}

Here, $(X, Y)$ is the union of $X$ and $Y$, and so $\mathit{support}(X=> Y) = \mathit{support}(Y => X)$.
\emph{Confidence of an association rule, $X => Y$, determines the probability that $Y$ will change in a commit operation provided that $X$ changed in that commit operation}. So the confidence of $X => Y$ can be determined  in the following way.

\begin{equation}
    \mathit{confidence}(X => Y) = \mathit{support}(X, Y) / \mathit{support}(X)
\end{equation}

In the technique GUI-RISP\textsubscript{TS}, we derive association rules between datasets and modules from pipelines and investigate those for providing suggestions regarding which intermediate state result from a workflow under progress should be stored.  

\textbf{Determining the supports and confidences of the  association rules obtained from the workflows.}

In the GUI-RISP\textsubscript{TS}, all the distinct association rules from all the previously executed workflows are determined, then the support and confidence of each of the rules are calculated in the following way.
 
\replaced[id=author1, remark={Addressed for the reviews of Round 2 Reviewer 3}]{
\textbf{Support of an association rule.} Support of an association rule is the number of times the rule (i.e., a subsequence of a workflow with a dataset) can be generated from the workflows. We used the same formula from the paper of Chakroborti et al. \cite{inproceedingsBigD18} for generating the association rules. The support of an association rule, $D \Rightarrow e_{ij}$ can be expressed formally in the following way.}{}

\begin{equation}
support (D \Rightarrow e_{ij}) = No.\,of\,times\,edge\,e_{ij}\,found\,in\,history.
\end{equation}

Here $e_{ij}  \epsilon E = (m_i, m_j), (1 \leq i \leq n, 1 \leq j \leq n, i \neq j)$.

\textbf{Confidence of an association rule.} We determine the confidence of the association rule $D \Rightarrow e_{ij}$ in the following way from its support value:

\begin{equation}
confidence (D \Rightarrow e_{ij}) = \frac{support (D \Rightarrow e_{ij})}{support(D)}
\end{equation}

Here, support(D) is the number of times input dataset D was used in all workflows.

\subsection{GUI-RISP\textsubscript{TS} Interaction Modelling}

We analyzed different situations in which users will utilize GUI-RISP\textsubscript{TS} in the user interface of a SWfMS. In the following paragraphs we also describe different situations of GUI-RISP\textsubscript{TS} in SciWorCS.

In this section of the study, we describe the possible workflow creation scenarios that inspired to design the user interface for the data recommendation technique. After investigating several SWfMSs and their workflows, we have figured out the following major composition scenarios.

\textbf{Composition scenarios.}
\begin{itemize}  
\item Creating a unique workflow by dragging and dropping tools. In this scenario, users may compose both small and large workflows from scratch, where sub workflows are unique to the system
\item Creating a workflow by dragging and dropping tools. In this case,  sub-workflows could be in the system from previous executions of different users with proper access control
\item Importing an existing workflow in the system from a saved one or a shared source. In this case, sub workflows could be in the system from previous executions of different users with proper access control.
\end{itemize}

Also, to increase the efficiency of the GUI-RISP\textsubscript{TS} selection process, we have considered the workflow change scenarios of workflow after composing the workflow in a SWfMS.

\textbf{Change scenarios.}
\begin{itemize}
\item Users could change modules, tune modules, delete modules from a composed workflow. 
Even user may want to save the modified workflow in the system
\item Parameter configurations could be happened in a module 
\item Tool state could be changed for a workflow.
\end{itemize}

By considering the above scenarios, recommendations are arranged on the SWfMS's interface.


\textbf{Recommendations.} 
\begin{itemize}
\item If a subset of modules is unique in a workflow then users will not get any recommendation to use any datasets rather a recommendation for saving datasets may appear on the interface
\item If a subset of modules can be found in previous executions, then users will get a recommendation for the subset if datasets are available otherwise a recommendation for saving datasets will appear on the interface
\item A message or recommendation list for a subsequence appears after adding a new module and connecting it to a previous one
\item Recommendation can be changed based on CRUD (i.e., create, read, update and delete) operations in a workflow
\item Recommendation can be updated after each parameter tuning, and configuration
\item Recommendations are based on tool states not the ids of tools.  
\item  Rather than creating a workflow from scratch if users choose to import it from existing ones, a dedicated button can be used to check the available intermediate data for subsequences
\item If the data loading time is more than the execution time for a module, users are warned with a message while using the recommendations
\item Required time to execute a module from previous execution is presented on the module for better understandability (i.e, knowing about how expensive a module is). 
\end{itemize}

\subsection{System Design}

A Job Manager of SciWorCS is responsible for executing the modules of an assembled workflow based on their dependencies and implementations (i.e., modules could be executed in a single server or a cluster). GUI-RISP\textsubscript{TS} acts alongside the job manager to suggest intermediate datasets automatically. Besides, users can manually check and load intermediate datasets of previously executed workflows while composing and executing a new workflow for reusing available datasets by the necessity of the new workflow execution using the GUI-RISP\textsubscript{TS}. That means when users won't like the recommendation, they might use the manual option to choose the intermediate states. One of the reasons for giving users multiple options is the required domain expertise in the workflow building process. If a user gets to know about the details (i.e., parameter settings) of the different suggestions, a suitable intermediate data can be chosen for her/his job. Since parameter configuration can change the intermediate data, domain expertise and user involvement are essential in the workflow building process. It is also possible to have suggestions with parameters that do not change the processed data for a workflow subsequence. Such as parameters for speeding up a module with more cores or other resources. In that case, the user might select any of the suggestions but need to know the parameter set. The design of the recommendation interface has emerged with such requirements. SciWorCS also provides real-time status monitoring, data mapping, and parameter setting mechanisms for each module individually to enable more modularized control. The Job Manager further performs the storing procedure of intermediate data and other necessary data by getting the recommendation from GUI-RISP\textsubscript{TS}. All of the execution records and configuration information of modules in a workflow are stored in a CouchDB server. GUI-RISP\textsubscript{TS} mainly associates the execution sequences and parameter information of previously executed workflows for recommending intermediate data while composing workflows in the SWfMS.

\subsubsection{Rules}
Figure \ref{UI} shows the interactive composition window of SciWorCS to assemble and execute workflows. Modules of a composed workflow could be executed in the Linux server as local scripts or in the Spark cluster as distributed jobs. Both distributed and local modules can be dragged and dropped from the left side toolbox panel (labeled as L) to the main container of workflows (labeled as MCW). After placing the required modules from the panel L for a workflow, users need to connect them with links to resolve their conditions and dependencies. The right side panel (labeled as R) in Figure \ref{UI} shows the available input datasets and output produced by workflows in the SWfMS. Datasets and parameters could be set for a module by double-clicking the module on a pop-up window (labeled as MPW). Each module in a workflow has a particular status label for the recommendation technique, GUI-RISP\textsubscript{TS} (labeled as GUI-RISP). Initially, this status for each module is `Not Checked' but after a certain time of assembling, checking or executing a module the status could be `Checked Not Found', `Checked Found', or `Load Data'. With the status, `Load Data', the SWfMS enables a button for loading intermediate data for an assembled workflow. The GUI version of RISP, where parameter matching and time are considered for giving multiple options to select intermediate data is placed below this status label as a List (labeled as PM and ET). By clicking an item from the list, a user can serve the same procedure of loading intermediate data for a workflow.

\subsubsection{Suggestion}


\begin{figure}\centering
\begin{tikzpicture}[zoomboxarray,
    zoomboxes below,
    zoomboxarray columns=3,
    zoomboxarray rows=1,
    connect zoomboxes,
    zoombox paths/.append style={ultra thick, orange}]
    \node [image node] { \includegraphics[width=1\textwidth]{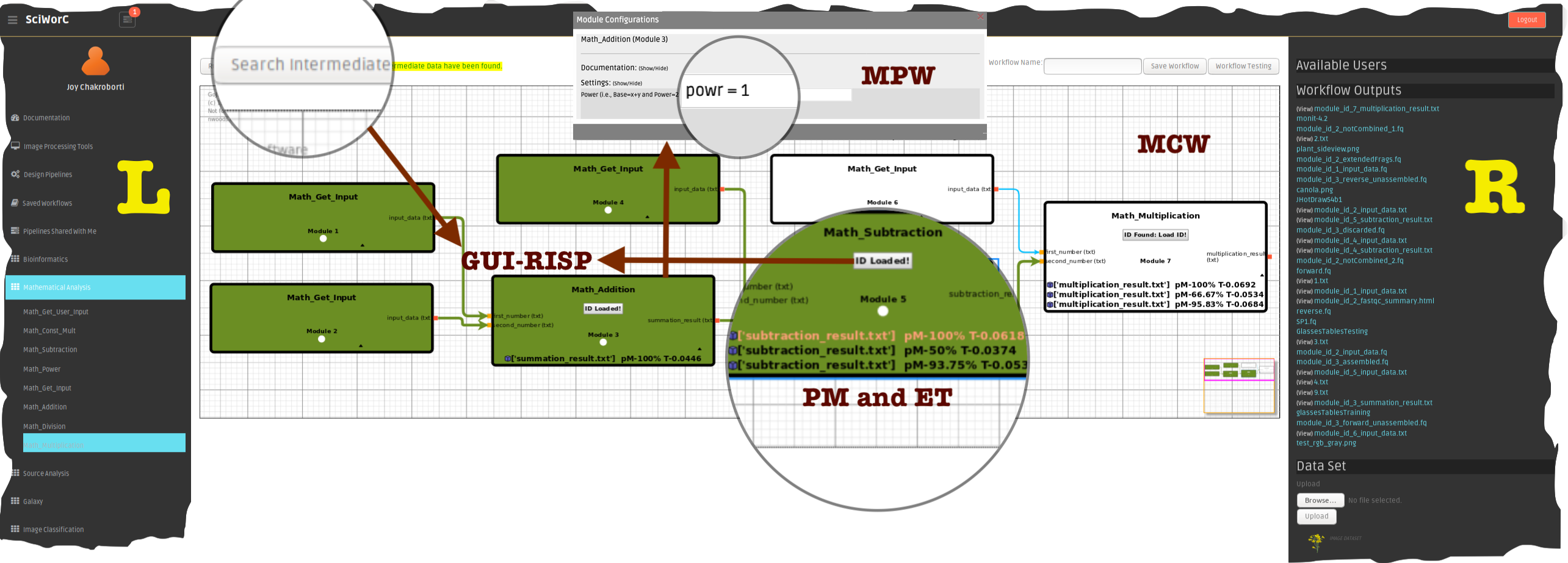} };
    \zoombox[magnification=2.1]{0.43, 0.80}
    \zoombox[magnification=1.8]{0.56,0.39}
    \zoombox[magnification=5]{0.777,0.45}
\end{tikzpicture}
\caption{(a) -User interface for building workflows. Panel L represents the available tools from the tool server. Panel R represents the available input and output data. The label GUI-RISP indicates the buttons to see intermediate data. MPW - module popup window is used for parameter configuration. PM and ET refer to the parameter matching percentage and estimated time for module execution, respectively. MCW - the main container window is used for whole workflow composition. (b) - Left - Parameter configuration, Middle - Action button and info, Right - Parameter matching percentage and estimated time.}
\label{UI}
\end{figure}

\replaced[id=author1, remark={Addressed for the reviews of Round 2 Reviewer 3}]{
Figure \ref{UI} also shows a workflow composition process in SciWorCS with the help of our proposed data recommendation technique, GUI-RISP\textsubscript{TS}. Following the assembling process for a desired workflow, the `Search' option could be used to know the existing relevant intermediate data that were stored by the recommendation of GUI-RISP\textsubscript{TS}. After designing a particular subsequence of a workflow, if a user clicks on the search button, the intermediate data loading buttons are displayed for the corresponding modules of the sequence. Module status label also displays the information for the possibility of loading and executing workflows with previously-stored intermediate data. In the figure, green-colored modules and links represent that the intermediate data are loaded up to the colored point, and there is no need of executions of those modules for the depicted workflow. In this case, module subtraction (i.e., module 5) had the option for selecting intermediate data, and the data loaded up to this module execution point. By observing the parameter matching information from the options of the GUI-RISP\textsubscript{TS} and selecting an intermediate dataset also serves the same fashion to execute a workflow in the Scientific Workflow Management System.}{}

\section{Implementation Details}
\label{ImplementationDetails}

SciWorCS \cite{10.1145/3331151} is built with Flask (A Python Microframework) to provide a web-based solution for the end-users. Various resource management techniques such as HDFS, CouchDB, UnixFileSystem are incorporated to enable interoperability among the used cutting edge technologies (i.e., Yarn, Spark, Python Subprocess and so on) in SciWorCS for our data management scheme. The core architecture of SciWorCS is implemented in Python 2, and the latest web technologies (i. e., JavaScript, GoJS, and so on) are used to provide the interactive user experience. The core implementation is hosted on a Linux server where all of the local scripts of modules are executed. A distributed Spark cluster (i.e., a Compute Canada Cluster of five-node, 40 cores and total 
of 200GB RAM)  and a CouchDB server are also associated with the SWfMS to submit distributed jobs and update log information from the web server by using their respective APIs.

\section{Experimental Studies and Results}
\label{ExperimentalStudiesandResults}

\replaced[id=author1, remark={Addressed for the reviews of Round 2 Reviewer 1, 5}]{Our experimental studies are divided into three sections. The three main studies are considered to include two domains and one user behavior experiments for GUI-RISP\textsubscript{TS}. The two domains are included for their popularity in the scientific workflow community. In addition, the domains (i.e., Image processing and Bioinformatics) are also important for the SciWorCS, as it is being used for the plant phenotyping and genotyping research in a University. As intermediate data are domain-dependent, their identifiers, as well as modules are unique in the system for each of the domains; they are kept unique every time they are generated with different configurations in the system. User behavior study is important to see the interest of users for the GUI-RISP\textsubscript{TS} in a SWfMS, which is driven by an eye-tracking device.}{}

Following hypotheses are considered before investigating the experimental results to answer our research questions in various studies:

\textbf{H1:} \textbf{GUI-RISP\textsubscript{TS} will not increase the overhead of SciWorCS}. \replaced[id=author1, remark={Addressed for the reviews of Round 2 Reviewer 1, 3}]{The reason behind this hypothesis is the possibility of minimal requests/responses from workflows for skipped-modules. If GUI-RISP\textsubscript{TS} can suggest intermediate data while assembling workflows, the number of module operations and status checking will be minimized in executions. Skipping some modules will ultimately not increase the overhead of using GUI-RISP\textsubscript{TS} in the SciWorCs. Verification of this hypothesis will answer our \textbf{RQ1:} How much performance overhead GUI-RISP\textsubscript{TS} adds in the system?}{}

\textbf{H2:} \textbf{User would use the intermediate data while executing workflows if the opportunity is given to them.} \replaced[id=author1, remark={Addressed for the reviews of Round 2 Reviewer 1, 3}]{We hypothesize that the user will use intermediate-data notably regarding the available intermediate-data. Here, we want to explore how much intermediate data are used in the SWfMS from available intermediate data. What are the percentages of using intermediate data with respect to the modules of the workflows. \textbf{RQ4} can be answered with the experimental results of this hypothesis.  How do users perceive intermediate states in composing workflows? This question is related to the usage of intermediate data, and their availability in the experiments can be used to answer the question.}{} 

\textbf{H3:} \textbf{GUI-RISP\textsubscript{TS} will help to compose workflows efficiently. We hypothesize modules' execution-time would be reduced if users use intermediate data}. \replaced[id=author1, remark={Addressed for the reviews of Round 2 Reviewer 1, 3}]{The reason behind this hypothesis is the introduction of a new technique in a scientific workflow management system, where users may need to provide their choice in assembling time. We hypothesized that users' unwanted delays might not affect the faster execution of workflows for skipping modules. So the combination of assembling and execution performance will be improved for using GUI-RISP\textsubscript{TS} in SciWorCS. This hypothesis test will also show us how we can use intermediate data in workflows. Results from this hypothesis can be used to answer our \textbf{RQ3:} Can we compose workflows while specifying intermediate states?}{}

Before going to the three main studies, we explored the load implication of the GUI-RISP\textsubscript{TS} in a SWfMS to answer our research question \textbf{RQ1}. That is, How much performance overhead GUI-RISP\textsubscript{TS} adds in the system? For assessing the workload of the GUI-RISP\textsubscript{TS} in SciWorCS, we have done the load testing, where we considered the SWfMS with the GUI-RISP\textsubscript{TS} and without the GUI-RISP\textsubscript{TS}. In our evaluation,  to avoid biases and unintentional delays from users, we used Apache Jmeter \cite{halili2008apache} and Google Analytics \cite{clifton2012advanced} for assembling workflows automatically and keeping logs with sampling. First, we recorded workflow compositions and executions of the two scenarios (i.e., the composition without GUI-RISP\textsubscript{TS} and with GUI-RISP\textsubscript{TS}) in the SWfMS using the Script Recorder of the Jmeter. Then we ran each of the scripts from Jmeter for 100 times to report its performance. Table \ref{table:overallload} compares various performance matrix of both of the scenarios for the SWfMS. In the table, we can see that using the GUI-RISP\textsubscript{TS} in SciWorCS does not change so much in average response time for various requests from our web application. Besides, the throughput (i.e., request-response completion/sec) increase with the GUI-RISP\textsubscript{TS}, and this is for the reason that only a very few requests and responses are needed in a workflow composition using the GUI-RISP\textsubscript{TS} in SciWorCS (i.e., 56\% fewer requests). \added[id=author1, remark={Addressed for the reviews of Reviewer 2}]{This supports our hypothesis \textbf{H1}; that is, GUI-RISP\textsubscript{TS} will not increase any overhead in the SciWorCS. So, \textbf{RQ1} can be answered partially that system-level performance will not be affected by the  GUI-RISP\textsubscript{TS}.}

\begin{table}
\tiny
\centering
\caption{Overall load of the SWfMS}
\begin{tabular*}{\columnwidth}{ @{\extracolsep{\fill}}|r|r|r| } 
\hline
\thead{Measures} & \thead{Composition without GUI-RISP\textsubscript{TS}} &  \thead{Composition with GUI-RISP\textsubscript{TS}} \\
\hline
Requests & 743 & 321 \\ 
Average Time  & 210.3(Sec) & 200.5(Sec)\\ 
Throughput/Sec & 3.5 & 5\\ 
Received KB/sec & 7.32 & 19.88\\ 
Sent KB/sec & 4.63 & 7.77\\ 
\hline
\end{tabular*}
\label{table:overallload}
\end{table}

To answer the \textbf{RQ1} fully, here, we inquire the overall end-user experience (i.e., user-level performance test) of the GUI-RISP\textsubscript{TS} and to do this, we followed the application performance monitoring (APM) strategy. By following the strategy, we mainly assess the overall user experience of GUI-RISP\textsubscript{TS} in SciWorCS. \added[id=author1,remark={Addressed for the reviews of Reviewer 3}]{Although we used the NASA-TLX questionaries to evaluate a more comprehensive user experience, this user experience is used to quantify the user-level performance of the SciWorCS. The NASA-TLX tests the usability from users' perspective, and the Apdex uses system performance value to illustrate users' satisfaction with the system.}

To complete the answer of \textbf{RQ1} and present the overall performance we calculate the Apdex Standard \cite{sevcik2005defining}, the equation to calculate the standard is given below -

\begin{equation}
Apdext = (SatisfiedCount + (ToleratingCount/2))/TotalSamples
\end{equation}

We chose Apdex in our initial evaluation because the Apdex score gives overall satisfaction for a system from all users rather than the satisfaction of a few individuals from different categories \cite{sevcik2005defining}. In the scoring process of Apdex to categorize the responses from users for a system, we need to define a threshold time. In the scoring we use the default threshold value (0.5 sec) to divide the satisfied and unsatisfied responses. To calculate the average single request response time of workflows, we execute 18 workflows in SciWorCS with GUI-RISP\textsubscript{TS} and keep log of the average respones time of each request (except module operation requests). 

From our Apdex scoring, with the value 0.89, we can say that near 90\% users are satisfied by the performances of SciWorCS after GUI-RISP\textsubscript{TS} integration. Thus, \textbf{RQ1} can be answered with a positive statement for the GUI-RISP\textsubscript{TS} being used in the SWfMS. Moreover, the average response time (i.e., in Table \ref{table:overallload}) is not varying much after GUI-RISP\textsubscript{TS} integration, so components can normally interact with each other after GUI-RISP\textsubscript{TS} integration in the SWfMS, which  also answers a part of \textbf{RQ1} for assessing the overall performance of the SWfMS. Furthermore, by integrating the GUI-RISP\textsubscript{TS} in SciWorCS, a higher number of KB/sec processing (Table \ref{table:overallload}) between a client and the server show the increased throughput for the SWfMS. So, no component is making the overall performance poor rather improving the performance of the SWfMS using the GUI-RISP\textsubscript{TS}.

For the details user evaluation, we mainly cover two popular scientific workflow composition areas (i.e., Image Processing and Bio-informatics workflow compositions) and one user behavior analysis by assembling and executing workflows with/without GUI-RISP\textsubscript{TS} in SciWorCS. All three studies are discussed below to answer some other important research questions (i.e., \textbf{RQ3} - \textbf{RQ5}).

\subsection{Participants}
We selected participants who had some knowledge on workflows (i.e., business or scientific workflows) as this study investigates the effectiveness of a technique in a SWfMS.  We selected a total of 7 participants aged from 24 to 35 from a university, including both graduate and undergraduate students. The participants had an average age of 28, where 43\% are female, 57\% are male. \added[id=author1, remark={Addressed for the reviews of Reviewer 3}]{All of the studies are performed with the same 7 participants.  In each session, two client environments are provided for both systems (i.e., a SWfMS with GUI-RISP\textsubscript{TS} and without GUI-RISP\textsubscript{TS}) to compose the workflows randomly. The workflow composition is time-consuming for both assembling and parameter tuning. In Study 1 and Study 2, we considered a total of 18 (9+9 for both systems) workflows, and they are assembled randomly for each session. So, the randomness and selection of two areas diminish the sequence effects and other conditions of doing unbiased studies.}

\subsection{Study 1: Image Processing Workflow Composition}

In study 1, we want to explore the user satisfaction towards the performance of composing and executing data intensive workflows by the aided data of our technique (i.e., GUI-RISP\textsubscript{TS}). Image processing workflows are used in this study to comprehend the differences in composing workflows with our proposed technique from the typical composition in SciWorCS. Generally, work pattern and efficiency of workflow composition and execution are explored for both of the scenarios (i.e., composition with and without GUI-RISP\textsubscript{TS}) in the SWfMS to answer our \textbf{RQ3} and \textbf{RQ4}. That are, Can we compose workflows while specifying intermediate states? How do users perceive intermediate states in composing workflows?

\replaced[id=author1, remark={Addressed for the reviews of Round 2 Reviewer 1,3}]{
In workflows of image processing where input datasets are normally large in size, the processing time of modules can be lengthy, and their produced intermediate data can be large to manage. A reusability technique of data might be effective for such workflows, and in this study, we want to ensure the effectiveness of our technique with image processing workflows. In the case of large data analysis, dependencies and conditions in workflows need to be considered especially to solve a desired problem in minimum cost. By considering these dependencies of directed acyclic type, several recommendation techniques \cite{Koop2008VisComplete:Pipelines} \cite{Chinthaka2009CBRAssistant} \cite{Zhang2011Recommend-as-you-go:Reuse} \cite{Spjuth2015ExperiencesBioinformatics} of modules or services are introduced.  However, none of these studies are presented with the usefulness of the recommended modules or intermediate data for reusing in workflow design. Therefore, in this study, to explore the usability of efficient workflow design by reusing intermediate data in a SWfMS, we use five image classification workflows of data-intensive computations. Consequently, how efficient the recommendation technique for users to perform workflow assembling and executing, how much time is required to design and execute a workflow and so on are analyzed in this study. Additional factors such as effects of a length, complexity (highest degree in a workflow) and data availability have been considered in our second study to assess the proposed technique and explore future directions in the workflow building process. Hence, in this part of our evaluation, we answer the \textbf{RQ3} and \textbf{RQ4} using the time factors and the data usage of workflow assembling and executing.}{}

\subsubsection{Task and Stimulus}
In the study, as mentioned above to verify the effectiveness of enhancing performance and gaining time, we consider five workflows of image classification. Modules of these five workflows perform distributed processing in the SWfMS's Spark cluster. For the classification models, we choose five popular deep neural network architectures, i.e., \textit{(InceptionV3, Xception, ResNet50, VGG16, and VGG19)} to give the option of choosing a model in our model-fitting module of the five workflows. In total, the five workflows contain ten distinct modules that are implemented in SCiWorCS and compatible with the recommendation technique, GUI-RISP\textsubscript{TS}. Each of the five workflows is computationally expensive and solves a classification problem using a sequence of data preparation, feature extraction, model fitting, and analysis. To eliminate biases, users had the opportunity to choose the classification models arbitrarily to perform image classification on a given dataset using a workflow from the five workflows. The dataset for the classification problem is collected from the Plant Phenotyping and Imaging Research Centre (P2IRC) of University of Saskatchewan and has more than 3K images (2K for training, 1K for testing). \added[id=author1,remark={Addressed for the reviews of Reviewer 4}]{Images were previously labeled for our classification problems}. In the study sessions, workflow structures are printed and provided to the participants. Three of the five workflows using the five models are explained to them as model testing problems of image classification. Workflow design procedure in SciWorCS such as dragging and dropping, port linking, dependency making of modules are described to the participants before the main study. Besides, how to check, load and use available intermediate data in the SWfMS are demonstrated using demo workflows to the participants.

\subsubsection{Experimental Procedure and Data Collection}
All of the occurred event, and their timestamps are logged in the background by defining JavaScript Tracking Snippet of Google Analytics \cite{clifton2012advanced} while building workflows by the participants.  To capture valuable information of assembling and executing workflows from both of the scenarios (i.e., SciWorCS with and without the GUI-RISP\textsubscript{TS}) each event is tracked with \textit{hitType, eventCategory, eventAction, eventValue, and eventLabel}. Similarly, timing information are tracked with \textit{hitType, timingCategory, timingVar, timingValue, and timingLabel} using the JavaScript Snippets of Google Analytics. At the end of the study,  participants were asked to fill out the \textit{NASA-TLX} questionaries for analyzing the workload of executing image processing workflows in the SWfMS.

\subsubsection{Results}

\begin{figure}[htbp]
\begin{tikzpicture}
\begin{axis}[
    ybar stacked,
    width=1\textwidth,
    bar width=15pt,
    ymin=0,
    ymax=75,
    bar shift=-10pt,
    xticklabel style = {xshift=-10pt},
    nodes near coords,
    every node near coord/.append style={xshift=-3pt,yshift=10pt,rotate=90},
    enlargelimits=0.15,
    legend style={at={(0.5,-0.20)},
      anchor=north,legend columns=-1},
    ylabel={\#Time(secs) },
    symbolic x coords={Wf1, Wf2, Wf3, Wf4, Wf5},
    xtick=data,
    x tick label style={rotate=45,anchor=east},
    ]
\addplot[ybar, black, draw=black, fill=darkgray!00, postaction={pattern = crosshatch dots}] plot coordinates {
(Wf1,25.56) (Wf2,21.96) (Wf3,23.09) (Wf4,26.70) (Wf5,24.9)};
\addplot[ybar, black, draw=black, fill=darkgray!00] plot coordinates {
(Wf1,47.72) (Wf2,41.04) (Wf3,42.28) (Wf4,45.05) (Wf5,40.32)};
\legend{\strut Assembling Time,  \strut Execution Time}
\end{axis}
\begin{axis}[
    ybar stacked,
    width=1\textwidth,
    bar width=15pt,
    bar shift=10pt,
    ymin=0,
    ymax=75,
    xticklabel style = {xshift=10pt},
    nodes near coords,
    every node near coord/.append style={xshift=17pt,yshift=10pt,rotate=90},
    enlargelimits=0.15,
    legend style={at={(0.5,-0.32)},
      anchor=north,legend columns=-1},
    symbolic x coords={Wf1wtI, Wf2wtI, Wf3wtI, Wf4wtI, Wf5wtI},
    xtick=data,
    x tick label style={rotate=45,anchor=east},
    ]
    \addplot+[ybar, black, draw=black, fill=darkgray!30, postaction={pattern = crosshatch dots}] plot coordinates {
    (Wf1wtI,31.12) (Wf2wtI,29.09) (Wf3wtI,26.13) (Wf4wtI,30.10) (Wf5wtI,27)};
  \addplot+[ybar, black, draw=black, fill=darkgray!30] plot coordinates {
    (Wf1wtI,22.72) (Wf2wtI,19.12) (Wf3wtI,15.15) (Wf4wtI,10.12) (Wf5wtI,11.16)};
\legend{\strut Assembling Time(WtI), \strut Execution Time(WtI)}
\end{axis}
\end{tikzpicture}
\caption*{WtI - with intermediate data}
\caption{Required assembling and execution time in image processing workflows}
\label{a_e_time}
\end{figure}

\replaced[id=author1, remark={Addressed for the reviews of Round 2 Reviewer 3}]{
In the study of image processing workflow's performances analysis, users activities of building and executing workflows are traced separately. Average assembling and executing time from all the participants for all the five workflows are plotted in Figure \ref{a_e_time}.}{} In Figure \ref{a_e_time} gray-colored bars represent assembling and executing time of the five workflows with the proposed technique, GUI-RISP\textsubscript{TS}. Dots pattern in a bar denotes the assembling time, and the solid part depicts the execution time. In the figure, we can see that assembling time of workflows with the GUI-RISP\textsubscript{TS} is higher than the assembling time of workflows without the GUI-RISP\textsubscript{TS} (i.e., The dotted gray bars are higher than the dotted white bars). However, all of the five gray bars (i.e., aggregated dotted and solid, marked with ``WtI" (with intermediate) coords) of the five workflows are lower in height than their corresponding white bar of the same workflow. That means overall performance of all of the workflows with the proposed technique GUI-RISP\textsubscript{TS} is better than the performances of workflows without the technique in SciWorCS. The reason behind this performance is that participants had the option to select recommended intermediate data by the GUI-RISP\textsubscript{TS}, and they used intermediate data to skip some modules operations. Though the assembling time of workflows with GUI-RISP\textsubscript{TS} is higher than the assembling time of workflows without GUI-RISP\textsubscript{TS}, the extra time in assembling for the recommendation and data configuration is negligible compared to the overall performance for all of the workflows. It should be noted that If the workflows are small, usually with 2 to 3 modules, then the technique might negatively impact the performance. However, our manual investigation and data of my-Experiment \footnote{https://www.myexperiment.org/workflows} show that usually workflow size is far more than that. Thus the proposed technique can efficiently handle intermediate data to recommend for reuse in a real-life data-intensive workflow building environment. The above result confirms our hypothesis \textbf{H3}, that is GUI-RISP\textsubscript{TS} will help to compose workflows efficiently. Our \textbf{RQ3} can be answered that the image processing workflows can be composed by efficiently specifying intermediate data with the technique.

\begin{figure}[htbp]
\begin{tikzpicture}
\begin{axis}[
    width=1\textwidth,
    enlargelimits=0.15,
    legend style={at={(0.5,-0.15)},
      anchor=north,legend columns=-1},
    ylabel={No. of Intermediate States Used},
    bar width=7mm,
    y=14mm,
    symbolic x coords={kWf 1, kWf 2, Wf 3, Wf 4, kWf 5},
    xtick=data,
    x tick label style={rotate=45,anchor=east},
    nodes near coords align={vertical},
    ]
\addplot[ybar, nodes near coords, fill=black!10] 
    coordinates {(kWf 1,1) (kWf 2,2) (Wf 3,2) (Wf 4,1) (kWf 5,4)};
\addplot[draw=olive!100,ultra thick,smooth] 
    coordinates {(kWf 1,2) (kWf 2,2) (Wf 3,4) (Wf 4,2) (kWf 5,4)};
\legend{ Used Data, Available Intermediate Data}    
\end{axis}
\end{tikzpicture}
\caption*{kWf X - known workflow X,      Wf X - workflow X}
\caption{Use of intermediate states in image procesing workflows}
\label{u_o_intermediate}
\end{figure}

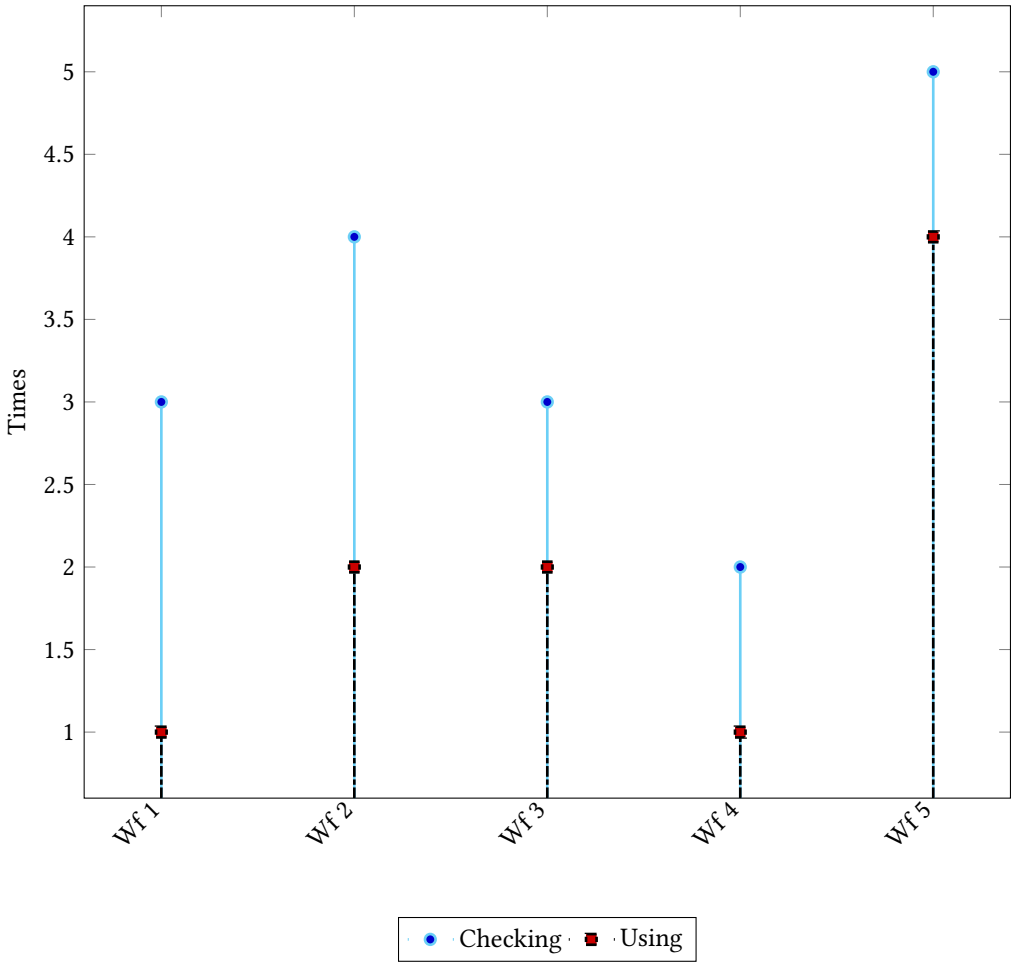
\begin{figure}[htbp]
\begin{tikzpicture} 
\begin{axis}[
    ylabel={Times},
    width=1\textwidth,
    symbolic x coords={Wf 1, Wf 2, Wf 3, Wf 4, Wf 5},
    xtick=data,
    x tick label style={rotate=45,anchor=east},
    legend style={at={(0.5,-0.15)}, anchor=north,legend columns=-1},
]
    \addplot+ [line width=1pt, draw=cyan!50,
        ycomb,
    ] coordinates {
        (Wf 1,3) (Wf 2,4) (Wf 3,3) (Wf 4,2) (Wf 5,5)
    };
    \addplot+ [line width=1pt,dash pattern=on 4pt off 1pt on 2pt off 1pt, draw=black!100,
        ycomb,
    ] coordinates {
        (Wf 1,1) (Wf 2,2) (Wf 3,2) (Wf 4,1) (Wf 5,4)
    };
\legend{ Checking, Using}  
\end{axis}
\end{tikzpicture}
\caption{Checking and Using Intermediate States in Image Processing Workflows}
\label{c_u_intermediate}
\end{figure}

Figure \ref{u_o_intermediate} depicts the data usage pattern among known and unknown workflows. ``Known workflow" to a user means the same workflow is executed by the same user previously. In our case, workflow 1, 2, and 5 (i.e., \textit{InceptionV3, Xception, and VGG19}) are executed by the participants prior to the main study while showing the demo of workflow building. We found that participants tend to use more intermediate data for known workflows rather than unknown workflows. In Figure \ref{u_o_intermediate}, around 87\% of available intermediate data are used for the known workflows, which are higher than the 50\% of two unknown workflows (N.B. all workflows had at least one intermediate data from previous executions). In addition, we found participants checked and used more intermediate data for computationally expensive modules (e.g., in our case, participant checked and used for the model fitting modules) of workflows rather than inexpensive modules. Traces of checking and using intermediate data in the workflows are plotted in Figure \ref{c_u_intermediate}. Figure \ref{c_u_intermediate} also represents the possibility of using intermediate data by checking the data from the users, and the results are more significant in all three known workflows (i.e., Workflow 1, 2 and 5) than the unknown 3 and 4. In this study, from the provided raw data and generated intermediate data, in total intermediate data and raw data showed the split of  83\% and 17\% usage while building workflows by the all participants. This information also showed that if we use the GUI-RISP\textsubscript{TS}, we can increase the reusability of a SWfMS. \added[id=author1, remark={Addressed for the reviews of Reviewer 2}]{This supports hypothesis \textbf{H2}, that is, user would use the intermediate data while executing workflows if the opportunity is given to them.} All of the findings in this study can be used to answer the \textbf{RQ4}, that is, users recognize the benefits of using intermediate states in workflow compositions, and up to this point, results show the GUI-RISP\textsubscript{TS} could be used efficiently for data-intensive workflows.

NASA-TLX responses from all of the participants for this study are plotted in Figure \ref{nasatlx_image}. In the figure, dark blue bars represent the average scores of all the participants for SciWorCS with the GUI-RISP\textsubscript{TS}. The gray bars represent the average scores of all the participants for SciWorCS without GUI-RISP\textsubscript{TS}. \added[id=author1, remark={Addressed for the reviews of Reviewer 2}]{The bar width represents the importance (i.e., weight) of the subclasses for our experiments, which is also collected from the participants of the studies at each session. Average of them are presented here for each subclass}. All of the subscales, such as  Mental Demand, Physical Demand, Temporal Demand, Frustration Level, Overall Performance, Effort of NASA-TLX are considered in this study with an overall score.
From Figure \ref{nasatlx_image}, it can be interpreted that on average, users require a low level of demands with intermediate data in all subclasses (i.e., the SWfMS with the GUI-RISP\textsubscript{TS}). Though assembling and executing time of workflows vary in the two systems. One reason could be, assembling workflow with the proposed technique in the SWfMS requires more demands as the participants needed to know more control of the data management. However, executing workflow with the GUI-RISP\textsubscript{TS}, users needed to wait less time, which could reduce the subclasses demands in the SWfMS. Overall, the aggregated effects of the two phases exhibit the low level of demands for the GUI-RISP\textsubscript{TS} system. \added[id=author1, remark={Addressed for the reviews of Reviewer 2}]{This result also supports hypothesis \textbf{H1} that is GUI-RISP\textsubscript{TS} will not increase the overhead of SciWorCS. Similarly, \textbf{RQ1} can be answered that the GUI-RISP\textsubscript{TS} does not add any negative performance implication in SciWorCS.}

\begin{figure}[htbp]
\pgfplotstableread{
  Scale    Mean  Weight
  MD      75    3
  PD      55    3
  TD      50    2
  FR      65    1
  OP      40    5
  EF      70    1
  Overall 58   2.5
  Dummy    0       0
}{\tlxweightedratings}
\pgfplotstableread{
  Scale    Mean2  Weight2
  MD      45    3
  PD      35    3
  TD      45    2
  FR      40    1
  OP      35    5
  EF      45    1
  Overall 41.67   2.5
  Dummy    0       0
}{\tlxweightedratingsInt}


\begin{tikzpicture}
  \begin{axis}[
    ymin=0,
    width=1\textwidth,
    ybar interval, xtick=data,
    ylabel={Ratings},
    legend style={at={(0.5,-0.15)},
      anchor=north,legend columns=-1},
    xticklabels={MD, PD, TD, FR, OP, EF, Overall},
  ]
    \addplot+[black,fill=black!50] table[x=Interval, y=Mean,
      create on use/Interval/.style={
        create col/expr={\pgfmathaccuma + \prevrow{Weight}}},
    ] {\tlxweightedratings};
        \addplot+[black, fill={rgb:red,1;green,2;blue,3}] table[x=Interval, y=Mean2,
      create on use/Interval/.style={
        create col/expr={\pgfmathaccuma + \prevrow{Weight2}}},
    ] {\tlxweightedratingsInt};
\legend{Without GUI-RISP\textsubscript{TS}, With GUI-RISP\textsubscript{TS}}    
  \end{axis}
\end{tikzpicture}
\caption*{MD-Mental Demand, PD-Physical Demand, TD-Temporal Demand, FR-Frustration, OP-Performance, EF-Effort, Overall- Overall Score}
\caption{NASA-TLX responses for image processing workflows}
\label{nasatlx_image}
\end{figure}


\subsection{Study 2: Bioinformatics Workflow Composition}

\replaced[id=author1, remark={Addressed for the reviews of Round 2 Reviewer 3}]{Scientific analysis of bioinformatics data is another popular domain that relies on SWfMSs for composing heterogeneous tools. In this study, we incorporate the bioinformatics workflow composition using the GUI-RISP\textsubscript{TS} and compare the composition with traditional workflow composition. In the previous study, we considered the same length workflows of heterogeneous technologies and large datasets. In this study, we consider composing micro-services or simple modules of bioinformatics (i.e., workflows are categorized in simple smaller, and complex longer) with comparatively smaller datasets. For the active bioinformatics communities, various micro-services and services are available and ready to use. To analyze bioinformatics data in a SWfMS, both simple and complex workflows are important for the availability of micro-services and services. Thus in this study, we conduct our experiment for both smaller and longer bioinformatics workflows to complete the answer of our \textbf{RQ3} and \textbf{RQ4}. That are, Can we compose workflows while specifying intermediate states? How do users perceive intermediate states in composing workflows?}{}

\subsubsection{Task and Stimulus}

In the previous study, we chose image processing workflows of distributed modules to process a massive amount of data. Contrast to that, here we choose four bioinformatics workflows of local-script modules and comparatively smaller datasets. First, two workflows are smaller in size and relatively easy to follow. The last two workflows are long in size and contain the first two workflows in its structure. In total, the four workflows contain four distinct modules such as \textit{Get\_input,  Pear, Fastqc, and Flash}. All of these modules are imported from Galaxy \cite{Afgan2016TheUpdate} to integrate into SciWorCS and compatible with the GUI-RISP\textsubscript{TS}. Each of the four workflows is not computationally expensive comparate to the previous study but complex in their structure for so many port mappings. The datasets for the bioinformatics problems are collected from the Plant Phenotyping and Imaging Research Centre (P2IRC) of University of Saskatchewan. The dataset has mainly two files, one is Forward FastQC, and another is Reverse FastQC.
\added[id=author1,remark={Addressed for the reviews of Reviewer 4}]{Both have 54,595 sequences with 300 sequence length and GC = 57\%.} In study sessions, workflow structures are printed and provided to the participants. All of the four workflows with some modules are explained to them as genome sequencing problems of bioinformatics. Likewise, previous study, workflow design procedure in the SWfMS such as dragging and dropping, port linking, dependency making of modules are described to the participants before the main study. Besides, how to check, load and use available intermediate data in SciWorCS are again demonstrated using demo bioinformatics workflows to the participants.

\subsubsection{Experimental Procedure and Data Collection}
Same as before all of the activities of the participants are logged and timestamped using JavaScript Tracking Snippet of Google Analytics while building workflows in the study season. Likewise before, we also capture the same information of assembling and executing workflows for both of the scenarios (i.e.,  SciWorCS with and without the GUI-RISP\textsubscript{TS}). Such as every event is tracked with \textit{hitType, eventCategory, eventAction, eventValue, and eventLabel} for all of the four workflows. Similarly, timing information is tracked with \textit{hitType, timingCategory, timingVar, timingValue and timingLabel} using the JavaScript Snippets of Google Analytics for all of the workflows. At the end of the main study, participant filled out the NASA-TLX questionnaire to evaluate the workload of executing bioinformatics workflows in the SWfMS.
 
\subsubsection{Results}

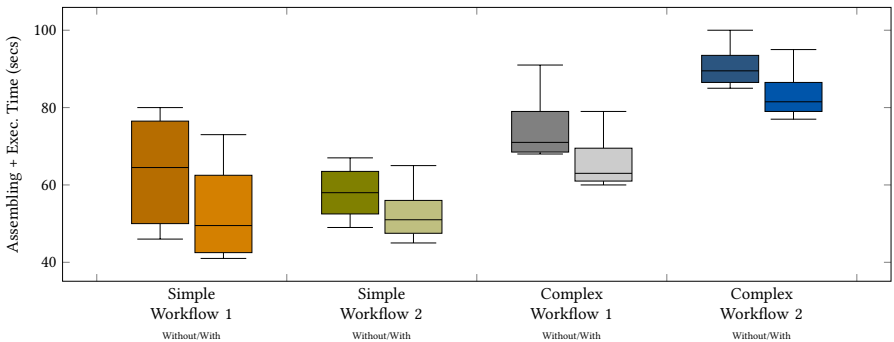
\begin{figure}[htbp]
\begin{tikzpicture}[scale=\textwidth/22cm]
\begin{axis}[
boxplot/draw direction=y,
ylabel={Assembling + Exec. Time (secs)},
boxplot={
    %
    draw position={1/3 + floor(\plotnumofactualtype/2) + 1/3*mod(\plotnumofactualtype,2)},
    %
    box extend=0.3,
},
x=4cm,
xtick={0,1,2,...,10},
x tick label as interval,
xticklabels={%
    {Simple Workflow 1\\{\tiny Without/With}},%
    {Simple Workflow 2\\{\tiny Without/With}},%
    {Complex Workflow 1\\{\tiny Without/With}},%
    {Complex Workflow 2\\{\tiny Without/With}},%
},
    x tick label style={
        text width=2.5cm,
        align=center
    },
]
\addplot[draw=black,fill={rgb:red,4;green,2;yellow,1}]
table[row sep=\\,y index=0] {data\\46\\54\\75\\78\\80\\};
\addplot[draw=black,fill={rgb:red,3;green,1;yellow,2}]
table[row sep=\\,y index=0] {data\\41\\44\\55\\70\\73\\};
\addplot[draw=black,fill=green!50!red]
table[row sep=\\,y index=0] {data\\49\\56\\60\\67\\86\\};
\addplot[draw=black,fill=green!50!red!50]
table[row sep=\\,y index=0] {data\\45\\50\\52\\60\\65\\};
\addplot[draw=black, fill=black!50]
table[row sep=\\,y index=0] {data\\68\\69\\73\\85\\91\\};
\addplot[draw=black, fill=black!20]
table[row sep=\\,y index=0] {data\\60\\62\\64\\75\\79\\};
\addplot[draw=black, fill={rgb:red,1;green,2;blue,3}]
table[row sep=\\,y index=0] {data\\85\\88\\91\\96\\100\\};
\addplot[draw=black, fill={rgb:red,0;green,0.5;blue,1}]
table[row sep=\\,y index=0] {data\\77\\81\\82\\91\\95\\};
\end{axis}
\end{tikzpicture}
\caption{Time differences among various types of bioinformatics workflow}
\label{s_c_time_bio}
\end{figure}

\begin{figure}[htbp]
\begin{tikzpicture}
\begin{axis}[
    width=1\textwidth,
    enlargelimits=0.15,
    legend style={at={(0.5,-0.15)},
      anchor=north,legend columns=-1},
    ylabel={No. of Int. States Used},
    bar width=7mm,
    y=18mm,
    symbolic x coords={sWf 1, sWf 2, lWf 1, lWf 2},
    xtick=data,
    x tick label style={rotate=45,anchor=east},
    nodes near coords align={vertical},
    ]
\addplot[ybar, nodes near coords, fill=black!10] 
    coordinates {(sWf 1,1) (sWf 2,1) (lWf 1,2) (lWf 2,3)};
\addplot[draw=olive!100,ultra thick,smooth] 
    coordinates {(sWf 1,1) (sWf 2,3) (lWf 1,2) (lWf 2,3)};
\legend{ Used Data, Available Intermediate Data}    
\end{axis}
\end{tikzpicture}
\caption*{sWf X - short workflow X, lWf X - long workflow X}
\caption{Use of intermediate states in bioinformatics workflows}
\label{u_o_intermediateS2}
\end{figure}

In this study of bioinformatics workflows, unlike the previous study assembling and executing time of each workflow is traced together. Average completion times of all of the four workflows are plotted in Figure \ref{s_c_time_bio}. The first box of each group in the box plot represents the total completion time of a workflow with the GUI-RISP\textsubscript{TS} (i.e., with intermediate data). The second box of each group represents workflow completion time of a workflow without the GUI-RISP\textsubscript{TS} (i.e., without intermediate data). First two groups in the figure depict completion time for the two simple workflows and last two depict completion time for the two complex workflows. \added[id=author1, remark={Addressed for the reviews of Reviewer 2}]{The reason behind using the box plot here is to point the long workflows constant completion time, which fluctuates more in small workflows, and intermediate data usage is not that helpful for small workflows.} From Figure \ref{s_c_time_bio}, it can be stated that for the complex workflows user can gain more time for the availability of intermediate data for a long sequence of modules than the simple/short workflows using the GUI-RISP\textsubscript{TS}. \added[id=author1, remark={Addressed for the reviews of Reviewer 2, 3 and 4}]{The above result also confirms our hypothesis \textbf{H3}, that is GUI-RISP\textsubscript{TS} will help to compose workflows efficiently in the real world as those are usually long (e.g., a MyExperiments workflow has on average 30 modules or more) in size. Our \textbf{RQ3} can be answered that the bioinformatics workflows can be composed by efficiently specifying intermediate data with the technique}. Availability of intermediate data and their usage can be clearly illustrated from Figure \ref{u_o_intermediateS2}. In the Figure, preferences of data usage while building workflows with GUI-RISP\textsubscript{TS} by the participants for both long and short workflows are represented. Similar to the previous study, this study also has workflows with at least one intermediate data from previous executions. 

\begin{figure}[htbp]
\begin{tikzpicture}
\begin{axis}[
    width=1\textwidth,
    ybar,
    enlargelimits=0.15,
    legend style={at={(0.5,-0.15)},
      anchor=north,legend columns=-1},
    ylabel={\#No of Modules},
    symbolic x coords={Wf1,Wf2,Wf3,Wf4},
    xtick=data,
    nodes near coords,
    nodes near coords align={vertical},
    ]
\addplot[fill={rgb:red,4;green,2;yellow,1}] coordinates {(Wf1,2) (Wf2,2) (Wf3,3) (Wf4,4)};
\addplot[fill=green!50!red] coordinates {(Wf1,0) (Wf2,0) (Wf3,2) (Wf4,2)};
\addplot[fill=black!50] coordinates {(Wf1,1) (Wf2,2) (Wf3,2) (Wf4,2)};
\addplot[fill={rgb:red,1;green,2;blue,3}] coordinates {(Wf1,1) (Wf2,1) (Wf3,2) (Wf4,3)};
\legend{UnKnown, UnKnown Used, Known, Known Used}
\end{axis}
\end{tikzpicture}
\caption{Use of intermediate states in bioinformatics workflows}
\label{s_c_useKvU}
\end{figure}
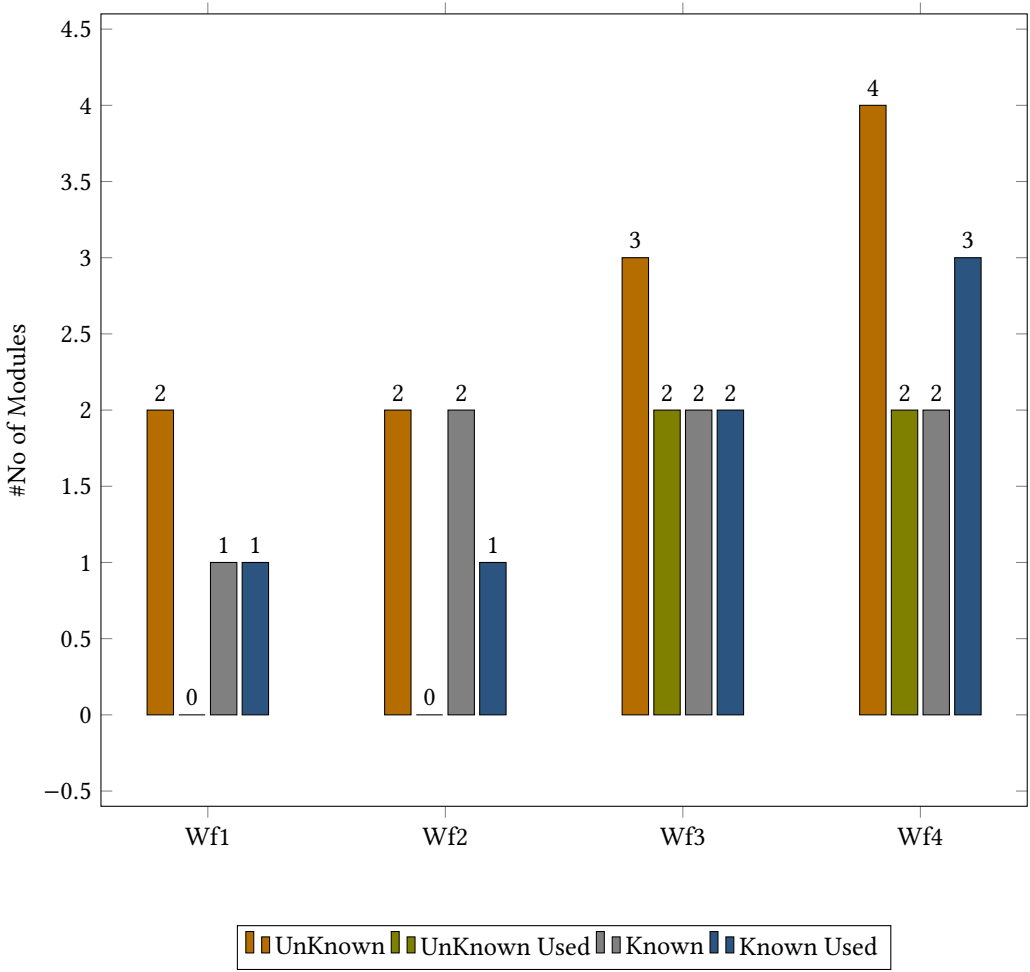

In the figure, we can see that for long workflows the participants used more intermediate data than the short workflows. In Figure \ref{s_c_useKvU}, we show the intermediate data usage for known and unknown modules. Known module means operation and relation of the modules in each workflow are explained to the participants prior to the study. For this study, known modules are Galaxy\_Pear and Galaxy\_Flash. From the figure, we can see that participants used more intermediate data to skip module operations for known modules than the unknown modules. In Figure \ref{s_c_useKvU}, 64\% of available intermediate data are used for the known modules, which are greater than the 36\% of unknown modules. This usage pattern could be from the reason that the participant felt comfortable to skip the operation of those modules by knowing that what they are skipping. So better understandability can improve the use of intermediate data, and in future, we will shift the SWfMS towards data-centric awareness system. In this study, from the provided raw data and generated intermediate data, in total intermediate data and raw data showed the split of  63 \% and 37 \% usage while building pipelines by all the participants. \added[id=author1, remark={Addressed for the reviews of Reviewer 2}]{This supports our hypothesis \textbf{H2}, that is, users would use the intermediate data while executing workflows if the opportunity is given to them, which is even more for the long workflows}. Experimental results of this study show that if we use the GUI-RISP\textsubscript{TS}, we can increase the reusability of a SWfMS for all types of workflows. The increased reusability and efficiency in the various types of workflows of this study complete the answer to our \textbf{RQ4} - users perceive intermediate states positively while composing workflows.  

\begin{figure}[htbp]
\pgfplotstableread{
  Scale    Mean  Weight
  MD      75    3
  PD      75    3
  TD      55    2
  FR      60    1
  OP      80    5
  EF      70    1
  Overall 69   2.5
  Dummy    0       0
}{\tlxweightedratings}
\pgfplotstableread{
  Scale    Mean2  Weight2
  MD      55    3
  PD      45    3
  TD      50    2
  FR      40    1
  OP      45    5
  EF      25    1
  Overall 41.33   2.5
  Dummy    0       0
}{\tlxweightedratingsInt}


\begin{tikzpicture}
  \begin{axis}[
    width=1\textwidth,
    ymin=0,
    ybar interval, xtick=data,
    ylabel={Ratings},
    legend style={at={(0.5,-0.15)},
      anchor=north,legend columns=-1},
    xticklabels={MD, PD, TD, FR, OP, EF, Overall},
  ]
    \addplot+[black,fill=black!50] table[x=Interval, y=Mean,
      create on use/Interval/.style={
        create col/expr={\pgfmathaccuma + \prevrow{Weight}}},
    ] {\tlxweightedratings};
        \addplot+[black, fill={rgb:red,1;green,2;blue,3}] table[x=Interval, y=Mean2,
      create on use/Interval/.style={
        create col/expr={\pgfmathaccuma + \prevrow{Weight2}}},
    ] {\tlxweightedratingsInt};
\legend{Without GUI-RISP\textsubscript{TS}, With GUI-RISP\textsubscript{TS}}    
  \end{axis}
\end{tikzpicture}
\caption*{MD-Mental Demand, PD-Physical Demand, TD-Temporal Demand, FR-Frustration, OP-Performance, EF-Effort, Overall- Overall Score}
\caption{NASA-TLX responses for bioinformatics workflows}
\label{nasatlx_bio}
\end{figure}

NASA-TLX responses from all of the participants for this study are also collected and plotted in Figure \ref{nasatlx_bio}. Same as the previous study, dark blue bars represent the average scores of all the participants for SciWorCS with the GUI-RISP\textsubscript{TS}. The gray bars represent the average scores of all the participants for the SWfMS without GUI-RISP\textsubscript{TS}. \added[id=author1, remark={Addressed for the reviews of Reviewer 2}]{The bar width represents the importance (i.e., weight) of the subclasses for our experiments, which is also collected from the participants of the studies at each session. Average of them are presented here for each subclass}. All of the subscales, such as  Mental Demand, Physical Demand, Temporal Demand, Frustration Level, Overall Performance, and Effort of NASA-TLX are also considered in this study with an overall score. From Figure \ref{nasatlx_bio}, it can be interpreted that on average, a user requires the low levels of subclass-demands in GUI-RISP\textsubscript{TS} system. This may be the reason that the completion time of workflows varies for their complexity and workflow completion with the proposed technique in the SWfMS for long workflows shows a higher gain in execution than the of short workflows. So, on average users are satisfied with the proposed technique in a SWfMS for its efficiency. 
 \added[id=author1, remark={Addressed for the reviews of Reviewer 2}]{Result from this bioinformatics case studies also supports hypothesis \textbf{H1} that is GUI-RISP\textsubscript{TS} will not increase the overhead of SciWorCS.}
With the response values from different subclasses of NASA-TLX in both of the studies, it can be interpreted that GUI-RISP\textsubscript{TS} does not increase workloads while composing workflows in SciWorCS. Thease results answer the \textbf{RQ1} of our experimental studies.

\subsection{Study 3: User Behavior in the Interface}
\label{study3}
 
\replaced[id=author1, remark={Addressed for the reviews of Round 2 Reviewer 3}]{In this study of user behavior analysis, we consider the time-spent and gaze-point matrics to explore the eye-tracking data of the participants from the workflow composition and execution. One of the main reasons to chose the eye-tracking technology for this study is to validate the user involvement on the intermediate data recommendation GUI (i.e., validation of the users' interest in intermediate data while composing workflows in SciWorCS). All of the workflows from the previous two studies are used to track user gaze-point and time duration by specifying different areas on the composing interface. Particularly, where participants are looking mostly and how much time they are spending in those areas are examined by Areas of Interest of Tobii Studio to answer the \textbf{RQ5}, that is, what are the major areas where participants are mostly involved while assembling and executing workflows with the GUI-RISP\textsubscript{TS}?}{}


Assembling and executing workflows in a SWfMS requires significant attention from users as there are some manual operations from users for both assembling and configuring the workflows. Thus, on an interface of workflow composition, we need to analyze where the gaze points are and where users are spending most of the time while composing workflows with the proposed management technique. Besides, we need to know whether the participants are really interested in intermediate data or is the GUI-RISP\textsubscript{TS} really helping them to compose workflows - consideration of these facts with our selected metrics should be adequate to answer the \textbf{RQ5}.

\subsubsection{Task and Stimulus} In this study of behavioral analysis, we did not consider any additional workflows. We collect both gaze data and time spent data (with Heatmap) for all of the workflows that are composed by the participant in Study 1 and 2. 

\subsubsection{Experimental Procedure and Data Collection}
Both the user behavioral data (i.e., gaze data and time spent data) are collected using a Tobii Eye-tracker while building workflows by the participants. The tracker was calibrated for each participant, and the participants maintained the recommended distance (i.e., between 18 to 40 inches) from the computer monitor. Important areas such as suggestion, and composition areas are divided into separate sections as `area of interest' for tracing eye-tracking data. Tobii Pro software solution is used for analyzing and representing the tracked data for the comparison.

\begin{table}[]
\tiny
\centering
\caption{Total Visit Duration}
\label{TotaVisitDuration}
\resizebox{\columnwidth}{!}{%
\begin{tabular}{|c|c|c|c|c|c|c|}
\hline
\multicolumn{7}{|c|}{www.p2irc-shipi.usask.ca/cvs}                                                                                                                                                                                                                                                                                                                                 
\\ \hline
\multirow{2}{*}{Summary Only} & \multicolumn{3}{c|}{Composition Area}                                                                                                                                    & \multicolumn{3}{c|}{Suggestion Area}                                                                                                                                     \\ \cline{2-7} 
                              & \begin{tabular}[c]{@{}c@{}}N\\ (Count)\end{tabular} & \begin{tabular}[c]{@{}c@{}}Mean\\ (Seconds)\end{tabular} & \begin{tabular}[c]{@{}c@{}}Sum\\ (Seconds)\end{tabular} & \begin{tabular}[c]{@{}c@{}}N\\ (Count)\end{tabular} & \begin{tabular}[c]{@{}c@{}}Mean\\ (Seconds)\end{tabular} & \begin{tabular}[c]{@{}c@{}}Sum\\ (Seconds)\end{tabular} \\ \hline
All Recordings                & \multicolumn{1}{r|}{7}                             & \multicolumn{1}{r|}{1539}                               & \multicolumn{1}{r|}{10773}                              & \multicolumn{1}{r|}{7}                             & \multicolumn{1}{r|}{596.30}                                & \multicolumn{1}{r|}{4174.1}                                \\ \hline
\end{tabular}%
}
\end{table}

\subsubsection{Results} 

\begin{figure}
  \includegraphics[width=1\textwidth]{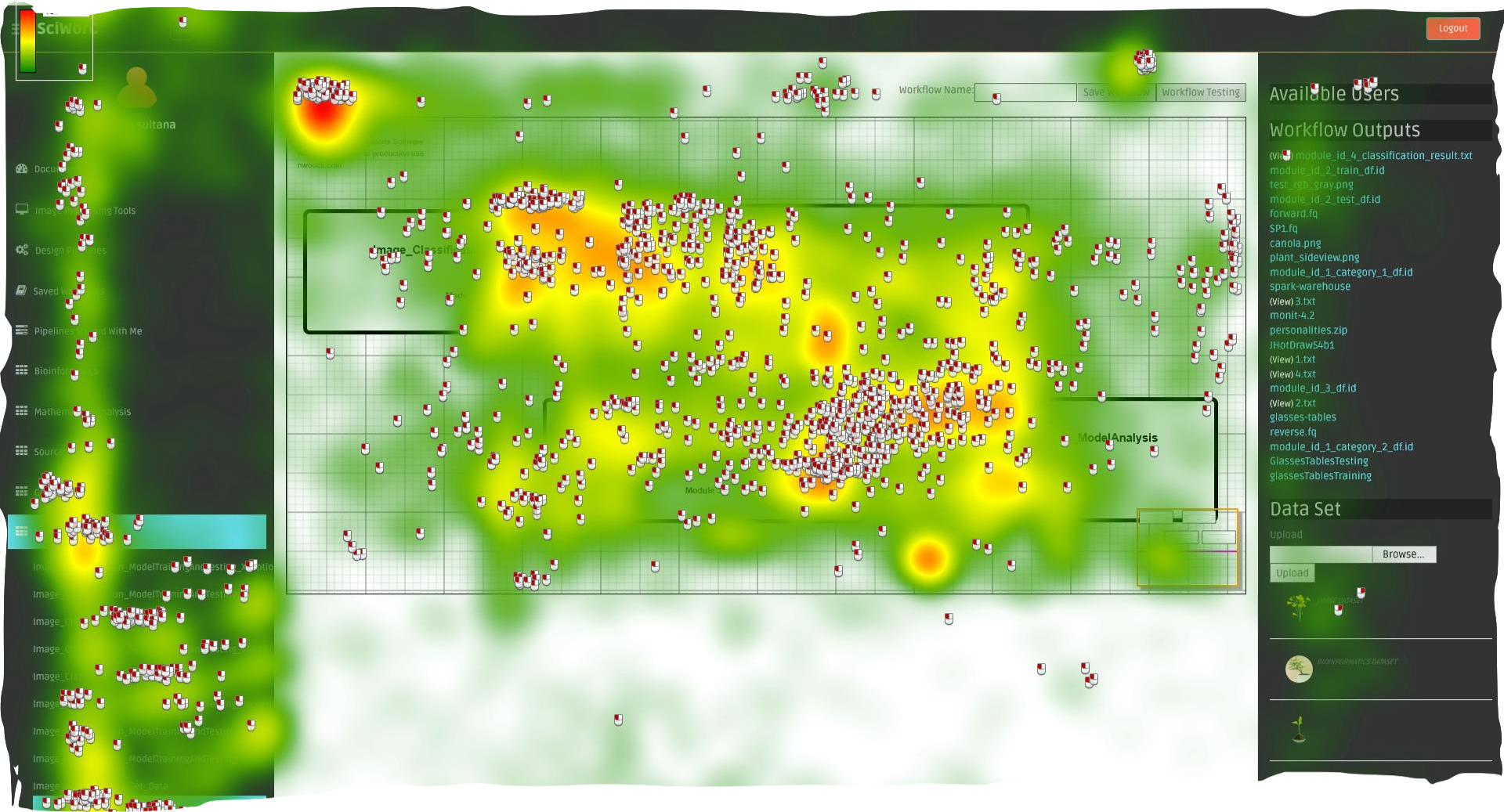}
  \caption{Heatmap of the interface after building workflows}
  \label{heatmap}
\end{figure}

Figure \ref{heatmap} is representing the heatmap of image processing workflows.  In the heatmap, we can see that users spent a considerable amount of time on modules after dragging and dropping them in the composition areas. Mostly on module 2 and 3, where users had the option to select intermediate data. So from the heatmap, it can be interpreted that users are interested in the suggestion of the recommendation technique.
Areas of our interest are divided into two regions, i.e., composition area and suggestion area. Table \ref{TotaVisitDuration} represents the average time spent data in these two areas separately. From the table, it can be seen that around 39\% time of total composition area time is spent on suggestion areas. This means with the facilities of using intermediate data with parameter and time information user tried to understand the suggested data with interest to use them. Result of this study answers our research question 5 (\textbf{RQ5}), i.e., participants are mostly involved in the suggestion areas while assembling and executing workflows and explore the potentiality of suggested data on the interface.

\section{Discussion and Future Directions}
\label{discussion_and_future_direction}

\replaced[id=author1, remark={Addressed for the reviews of Round 2 Reviewer 3}]{The results of the three studies indicate that users generally prefer intermediate data and can be benefitted from reusing while composing workflows in a SWfMS.}{} Our first research question \textbf{RQ1} is - How much performance overhead GUI-RISP\textsubscript{TS} adds in the system? Where our hypothesis \textbf{H1} was GUI-RISP\textsubscript{TS} will not increase any overhead in the SciWorCS. We found that both system-level and user-level performances of SciWorCS are not affected by the GUI-RISP\textsubscript{TS}. Similarly, the NASA-TLX study exhibits a low level of subclass demands for the GUI-RISP\textsubscript{TS} technique in SciWorCS for image processing and bioinformatics studies. We also found that our proposed technique can efficiently handle intermediate data to recommend for reuse in a real-life data-intensive workflow (i.e.,  image processing workflows) and long workflows (i.e., bioinformatics workflows )  building environment. This confirms our hypothesis \textbf{H3} - GUI-RISP\textsubscript{TS} will help to compose workflows efficiently. So,  \textbf{RQ3} is answered as that both image processing and bioinformatics workflows can be composed by efficiently specifying intermediate data with the technique.  We found our hypothesis \textbf{H2}  is valid from the data usage pattern experiments - users will use the intermediate data while executing workflows if the opportunity is given to them in both image processing and bioinformatics studies.  This answers \textbf{RQ4} - users can recognize the benefits of using intermediate states. Our last research question, i.e., \textbf{RQ5}, is answered from the data of the eye-tracking study. We found participants are mostly involved in the suggestion areas while assembling and executing workflows to explore the potentiality of suggested data on the interface. In the study of computationally expensive image processing workflows, we found that the participants prefer mostly to use intermediate data for computationally expensive modules. In the future, we plan to prioritize the expensive modules in SciWorCS to store their outcomes with the technique so that we can increase the reusability respect to users' usage patterns. Other significant findings that explored some research directions are also considered in this study to improve our recommendation system further. Especially the finding of users' usage preferences on intermediate data for known modules and known workflows of the two studies could be used to design a new architecture of the SWfMS where data associations with module and workflows will be expedited for efficiency. For instance, in real-life complex workflows are formed from a different domain, and many collaborators of different areas are involved in the workflow design process. In such a case, intermediate data management can help a collaborator design and execute a known part of a workflow. Other collaborators could reuse this part for the rest of the execution. This type of system is only possible if we can increase the data awareness in a SWfMS, and we are currently working on it to use intermediate data in a collaborative environment.  In our second study of bioinformatics workflows, we see that while assembling long workflows, users prefer to use more intermediate data. Since in real life workflow building process, sometimes 100 or more modules are involved so GUI-RISP\textsubscript{TS} can have a positive impact on SWfMS. Furthermore, each of the overall scores of NASA-TLX responses from both of the studies separately shows that after integrating the GUI-RISP\textsubscript{TS} with SciWorCS, participants experience less overhead. Moreover, the subclass's scores are less after integrating the GUI-RISP\textsubscript{TS} in the SWfMS. Hence we believe the technique of recommending intermediate data can increase the efficiency in a SWfMS without any shortcomings.

\section{Threat to Validity}
\label{ThreattoValidity}

We considered workflows of different but related domains (Image Processing and Bioinformatics) and one behavior test in our user study for analyzing the effectiveness of the GUI-RISP\textsubscript{TS} in SciWorCS. While evaluating with more domains could have made our findings more generalized, we see that the GUI-RISP\textsubscript{TS} shows effectiveness on workflows from both of the domains in different contexts.  Thus, we believe that our findings cannot be attributed to a chance. The proposed technique GUI-RISP\textsubscript{TS} can be considered an efficient technique for managing workflows. \added[id=author1,remark={Addressed for the reviews of Reviewer 4}]{Another threat of the study is the limited number of participants. However, in our experiments, we considered a total of 18 workflows for each session of a  subject.  Each session took almost two hours to compose 18 workflows in two systems from two domains. We believe the results from the above time-consuming studies support us in generalizing our findings in most cases.}
\section{Conclusion}
\label{ConclusionandFutureWorks}

To implement an efficient system for processing a large amount of data with numerous tools, proper data management in a system of workflow management is crucial. Here, we propose a technique of intermediate data recommendation for both storing and retrieving data while building workflows in a SWfMS. In addition, this study is intended to investigate case studies of a GUI version of the technique for comprehending users' behaviors and expectations in real-world workflow building. In our two case studies, from the provided raw data and generated intermediate data, in total intermediate data and raw data showed a split of 73\% and 27\% usage while building workflows by all participants. This usage trend and preference on intermediate data imply that the technique can fulfill users' expectations in most cases of efficient workflow execution. Additionally, we found that the technique is more useful for the long-running and complex workflows, so we believe our technique (i.e., GUI-RISP\textsubscript{TS}) has the potential to contribute to composing workflows with big data and heterogeneous tools. Moreover, using the NASA-TLX study, we found that demand in each subscale of NASA-TLX does not produce any negative impact while assembling and executing workflows in the SWfMS with the technique. Therefore, we believe the technique can be used in any SWfMS without any additional burden for introducing reusability and building workflow efficiently.




\begin{acks}
This work is supported in part by the Canada First Research Excellence Fund (CFREF) under the Global Institute for Food Security (GIFS).
\end{acks}

\bibliographystyle{ACM-Reference-Format}
\bibliography{main}




\end{document}